\documentclass[useAMS,usenatbib,fleqn]{mnras}
\usepackage{amsmath}
\usepackage{multirow}
\usepackage{graphicx}
\usepackage{color,soul}
\usepackage{epsfig}
\usepackage{amssymb,amsmath}
\usepackage{aas_macros}
\usepackage{fixltx2e}
\title[SMBH Assembly]{
Exploring SMBH Assembly with Semi-analytic Modelling
}

\author[Ricarte \& Natarajan]{Angelo Ricarte$^1$,
Priyamvada Natarajan$^1$ \\
$^1$ Department of Astronomy, Yale University, 52 Hillhouse Avenue, New Haven, CT 06511 \\
}

\date{\today}

\begin{document}
\pagerange{\pageref{firstpage}--\pageref{lastpage}} \pubyear{2017}
\maketitle

\begin{abstract}
We develop a semi-analytic model to explore different prescriptions of supermassive black hole (SMBH) fuelling.  This model utilises a merger-triggered burst mode in concert with two possible implementations of a long-lived steady mode for assembling the mass of the black hole in a galactic nucleus.  We improve modelling of the galaxy-halo connection in order to more realistically determine the evolution of a halo's velocity dispersion.  We use four model variants to explore a suite of observables:  the $M_\bullet-\sigma$ relation, mass functions of both the overall and broad-line quasar population, and luminosity functions as a function of redshift.  We find that ``downsizing'' is a natural consequence of our improved velocity dispersion mappings, and that high-mass SMBHs assemble earlier than low-mass SMBHs.  The burst mode of fuelling is sufficient to explain the assembly of SMBHs to $z=2$, but an additional steady mode is required to both assemble low-mass SMBHs and reproduce the low-redshift luminosity function.  We discuss in detail the trade-offs in matching various observables and the interconnected modelling components that govern them. As a result, we demonstrate the utility as well as the limitations of these semi-analytic techniques. 
\end{abstract}


\section{Introduction}
\label{sec:introduction}

Most massive galaxies in the universe harbour supermassive black holes (SMBHs) at their centres \citep{Kormendy&Richstone1995}.  Their existence has been confirmed via a range of observational techniques - gas, stellar, or maser dynamics \citep[see][for recent compilations]{Saglia+2016,vandenBosch2016}, including Sgr A* at the centre of our own galaxy \citep{Ghez+2008,Genzel+2010,Gillessen+2016}.  These studies have revealed correlations between SMBH mass and various host galaxy properties, the most notable of which is the $M_\bullet-\sigma$ relation, that relates the SMBH mass to the velocity dispersion of the inner regions of the host galaxy \citep{Ferrarese2002,Kormendy&Ho2013}.  These relations are likely the result of SMBH and host galaxy coevolution.

Yet it remains unknown how and when these relations are established and if they evolve with redshift.  In the context of the larger structure formation model within which black hole growth is embedded along with galaxy assembly -  the hierarchical, merger-driven model -  one explanation for this observed correlation is that SMBHs and their hosts grow in concert over cosmic time.  This could happen when a major merger funnels gas to the centre of the galaxy, triggering both an active galactic nucleus (AGN) and a central starburst \citep{DiMatteo+2005}.  One clue supporting this picture is the so-called ``active galactic nuclei (AGN) main sequence,'' a linear correlation between SMBH accretion rates and star formation rates inferred from SED fitting that was first observed by stacking star-forming galaxies \citep{Mullaney+2012}.  Others suggest that these relations could simply be the result of the central limit theorem: it is possible to create a correlation via many mergers of initially uncorrelated SMBHs and host galaxies \citep{Peng2007,Jahnke&Maccio2011}.  Observationally, it has been challenging to determine the precise nature of the link between mergers and AGN \citep{Mechtley+2016}.  It has been suggested that perhaps only the most luminous AGN are triggered via major mergers \citep{Treister+2012,Hong+2015}, although models of this too are currently debated \citep{Villforth+2017}.  

In order to trace how SMBHs grow, determining both their mass functions and luminosity functions as they assemble is paramount.  The local SMBH mass function serves as a boundary condition to the evolving population of SMBHs, while luminosity functions inform us about its rate of change over cosmic time.  Two broad classes of models have emerged using these constraints.  On one hand, empirical ``population synthesis'' models use the continuity equation to integrate the local mass function backwards using luminosity functions \citep{Small&Blandford1992,Haehnelt+1998,Merloni&Heinz2008,Tucci&Volonteri2017}.  On the other, semi-analytic models (SAMs), start with an initial population of high redshift  black hole ``seeds'' and attempt to match constraints by integrating forward with physical prescriptions for SMBH growth \citep{Volonteri+2003,Somerville+2008,Lacey+2016}. Both approaches have their strengths and limitations, and they have provided us with invaluable insights into modelling the SMBH population as a whole while incorporating empirical constraints.

Semi-analytic models have typically struggled to reproduce luminosity functions and mass functions simultaneously for actively growing SMBHs, particularly at the high mass/luminosity end.  \citet{Natarajan&Volonteri2012} compare the mass functions generated by their SAM to both empirical models of the overall population and broad-line quasar (BLQ) masses of the active population.  They find that the SAM predicts a dearth of high-mass AGN (from BLQ scaling relations) at high-redshift, but an overabundance of high-mass SMBHs (from galaxy scaling relations) at low-redshift.  In other words, SAMs need to rapidly establish a population of $>10^9 \ M_\odot$ SMBHs at redshifts of $z \gtrsim 5$, but avoid overgrowing them by $z=0$.  SAMs embedded in galaxy formation models are able to reproduce luminosity functions, but only by invoking an ad-hoc sub-Eddington cap on the accretion rate \citet{Hirschmann+2012}.  Related to this phenomenon is so-called AGN downsizing - the observation that the number densities of AGN with higher luminosities peak at higher redshift \citep{Hasinger+2005,Ueda+2014}.

In this paper, we develop a SAM with the aim of improving the match to bolometric luminosity functions and mass functions simultaneously.  In particular, we explore in detail the basic set of physical processes that are required to reproduce observables.  Like many previous works, this SAM takes a minimal approach to modelling galaxy evolution, instead focusing on the physics behind SMBH seeding and growth via accretion.  We explore a variety of modelling ingredients, to take the relevant astrophysical processes into account and select a few interesting sets of models that we discuss in detail.

In \S\ref{sec:otherSAMs}, we first review previous modelling work in this field.  In \S\ref{sec:modelling}, we describe the recipes utilised in the SAM, including two new key ingredients: the inclusion of a long-lived steady accretion mode and improved mapping of the galaxy-halo connection in modelling the dark matter haloes that harbour the assembling SMBHs in their galactic nuclei.  In \S\ref{sec:results}, we discuss how each of our models fares at reproducing the mass and luminosity functions over cosmic time.  In \S\ref{sec:discussion}, we discuss the limitations of this work and the effects of each of our modelling parameters.  In \S\ref{sec:conclusion}, we summarise our work and discuss prospects for future studies.  Throughout, we use the cosmological parameters of Planck 2016: $\Omega_m=0.31$, $\Omega_\Lambda = 0.69$, $\Omega_b = 0.049$, $\Omega_k=0$, $h = 0.68$, $\sigma_8 = 0.82$, and $n=0.97$ \citep{Planck2016}.  We use $\log$ to denote a logarithm of base 10, and $\ln$ to denote a logarithm of base $e$.

\section{Previous Modelling Efforts}
\label{sec:otherSAMs}

Semi-analytic models link the formation and growth of SMBHs to the evolving properties of their parent galaxies and/or dark matter haloes.  Some embed the growth of SMBHs within a full galaxy formation framework \citep{Kauffmann&Haehnelt2000,Somerville+2008,Bower+2010,Fanidakis+2011}.  In such models, the SMBH growth rate is often set to be proportional to the star formation rate in the bulge, with the possible addition of a long-lived, low accretion mode.  Other models strive to make minimal assumptions about galaxy properties, instead tying SMBH growth mainly to halo mergers \citep{Volonteri+2003,Volonteri+2013}.  This approach allows for improved halo mass resolution that enables the testing of different models for seeding, accretion, and other processes, such as the detailed dynamical evolution of binaries \citep{Volonteri+2008}.

Briefly, we summarise some of the successes and challenges for these models.

\begin{itemize}

\item Discriminating Between Seeding Models:  Cosmological simulations normally lack the dynamic range in mass necessary to resolve SMBH seed formation, the locations of seeding sites and the subsequent accretion onto black holes with masses much less than $10^6 \ M_\odot$.  Hence, semi-analytic models are more efficient for studying the consequences of different initial seeding models.  Light seeds, such as those formed from Pop III remnants, result in larger SMBH occupation fractions at $z=0$ compared to heavy seeds, and the $M_\bullet-\sigma$ relation evolves with redshift in a much different manner \citep{Volonteri+2008}.  Depending on the accretion prescriptions employed, they may also struggle to reproduce the high-luminosity end of the luminosity function at $z \sim 6$.

\item AGN-Triggered Growth:  While initial modelling efforts included only merger-driven accretion \citep{Volonteri+2003,Somerville+2008}, newer models have included secular accretion modes as well.  These studies agree that most local Seyferts can be fuelled by disk instabilities, but galaxy interactions are likely required to trigger the most luminous AGN with extremely high accretion rates \citep{Hirschmann+2012,Menci+2014,Gatti+2016}.  

\item High-redshift Quasars:  Nearly a hundred luminous quasars have been found at $z \sim 6$ \citep{Fan+2003} and the high-luminosity end of the luminosity function has been constrained \citep{Jiang+2016}.  Modelling the assembly of these ``monsters'' usually involves computing merger trees of highly biased $\sim 10^{13} \ M_\odot$ halos at $z=6$.  Studies have revealed that growing SMBHs to these extreme masses is possible if super-Eddington accretion rates are allowed \citep{Pezzulli+2016}.  Such rapid growth episodes can compensate for the low duty cycle of these accretion events \citep{Pezzulli+2017,Prieto+2017}, as well as the loss of some SMBHs due to gravitational wave recoil \citep{Volonteri&Rees2006}.

\item Assembling the Most Massive Black Holes:  The assembly of the most massive SMBHs ($M_\bullet \gtrsim 10^9 \ M_\odot$) at every cosmic epoch has been difficult to model successfully.  When comparing to mass functions derived from population synthesis models or broad-line quasars detected by the Sloan Digital Sky Survey (SDSS) for redshifts $1.5 < z < 4.5$, \citet{Natarajan&Volonteri2012} found that that SMBH SAMs systematically underproduced high-mass SMBHs at high redshift, but overproduced them at low redshift.  To solve this problem, new accretion modes have been proposed.  In particular, it is well-established that secular, non-merger-driven modes of AGN triggering are necessary, especially at low-redshift \citep{Hirschmann+2012}.  The exact nature of this secular mode remains an open question.

\item Evolution of SMBH Spin:  Included here for completeness, SAMs are an effective tool for predicting the spin-evolution of SMBHs.  Spin is thought to be relevant for a variety of mechanisms, including the radiative efficiency of the accretion disk \citep{Sadowski2009}; the powering of jets \citep{Blandford&Znajek1977}; and the emission of gravitational waves during SMBH mergers \citep{Lousto+2010}.  Broadly speaking, SMBHs are expected to spin up during periods of high Eddington ratio, when the accretion disk can align with the spin of the SMBH \citep{Natarajan&Pringle1998}, and randomise as the result of secular accretion or SMBH mergers \citep{Volonteri+2005,Volonteri+2013,Fanidakis+2011}.  However, these predictions are highly sensitive to ad-hoc assumptions about the timescale of the alignment of the accretion disk relative to the SMBH spin axis.
\end{itemize}

\section{Our Semi-analytic Model}
\label{sec:modelling}

Like many other previous works, our SAM uses a Monte Carlo merger tree as its backbone.  The seeding and growth of SMBHs are linked to the merger histories of their hosts.  Scaling relations are used to determine galaxy properties, the most important of which is its central velocity dispersion.

\subsection{Monte Carlo Merger Trees}

We use standard techniques to generate merger trees within the $\Lambda$CDM paradigm.  For each realisation of the universe, we generate 15 merger trees of 20 host halo masses sampled logarithmically between $10^{11}$ and $10^{15} \ M_\odot$ at $z=0$.  820 time steps are taken between $z=20$ and $z=0$, logarithmically sampling the expansion factor $a = 1/1+z$.  Our SAM utilises the binary merger tree code of \citet{Parkinson+2008}.  Specifically tuned to the Millennium simulation \citep{Springel+2005}, this code has been shown to most faithfully reproduce halo merger rates compared to other algorithms \citep{Jiang&vandenBosch2014}.  

We make one adjustment to the code in order to resolve the halos hosting SMBH seeds while limiting the size of the merger tree.  Our solution is to enable a redshift-dependent resolution scheme, as has been done in previous work \citep{Volonteri+2003}.  At redshift $z$, the resolution of the merger tree is set to 
\begin{equation}
M_\mathrm{res} = f_\mathrm{res} M_0 \left( \frac{1+z}{1+z_0} \right)^{-\zeta}.
\end{equation}
\noindent where $M_0$ is the host dark matter halo mass, $z_0$ is the host halo redshift, and $f_\mathrm{res}$ is the fractional resolution at $z=z_0$.  The redshift evolution slope $\zeta$ is given by
\begin{equation}
\zeta = \frac{\log(f_\mathrm{res} M_0 / M_\mathrm{min})}{\log[(1+z_\mathrm{max})/(1+z_0)]}.
\end{equation}
\noindent where $M_\mathrm{min}$ is the minimum halo mass resolved at the maximum redshift $z_\mathrm{max}$.  Since we wish to resolve the halos where SMBH seeding takes place, we set $f_\mathrm{res} \equiv 10^{-3}$, $M_\mathrm{min} \equiv 5 \times 10^{6} \ M_\odot$, and $z_\mathrm{max} \equiv 20$.

\begin{figure}
   \centering
   \includegraphics[width=0.45\textwidth]{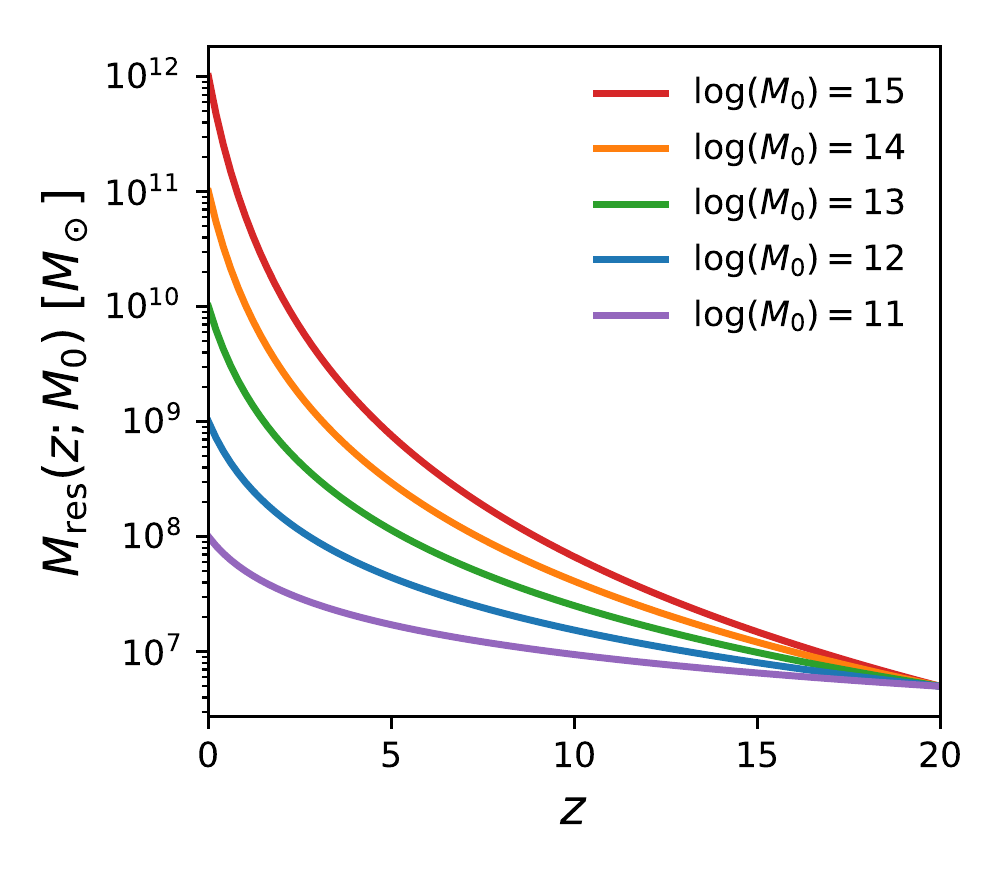}
   \caption{Redshift evolution of the mass resolution scheme.  As we note, the merger trees are designed to reach $5 \times 10^6 \ M_\odot$ at $z=20$.}
   \label{fig:massResolution}
\end{figure}

In figure \ref{fig:massResolution}, we plot the mass resolution of our SAM as a function of redshift for 5 different host halo masses at $z=0$.  The resolution continuously slopes towards the same value at $z=20$ for all masses.

\subsection{Halo Properties}

\subsubsection{Halo Spin Parameters}

In models wherein a massive initial seed forms in a pristine, gas-rich collapsed dark matter halo at high redshift - the direct collapse black hole (DCBH) channel - there exists an initial correlation between halo properties and that of the SMBH.  Halo spin parameters are drawn randomly from a log-normal distribution with a mean of $\bar{\lambda}=0.05$ and a standard deviation of $0.5$ \citep{Warren+1992}.  The evolution of the spin parameter is not relevant in our model, as only halos without progenitors are allowed to host seeds.

\subsubsection{Virial Temperature}

The virial temperature of a dark matter halo is directly related to its velocity dispersion, a property that is key in our modelling of the putative hosts of SMBHs.
Necessary once again for our seeding model, virial temperatures are set to those appropriate for an isothermal sphere \citep{Mo+2010}.  This is given by

\begin{equation}
T_\mathrm{vir} = 3.6 \times 10^5 \, K \left( \frac{V_{c,\mathrm{iso}}}{100 \, \mathrm{km}\,\mathrm{s}^{-1}}\right)^2.
\end{equation}

Following the spherical collapse model, a halo of mass $M_h$ at redshift $z$ has a circular velocity given by \citep{Barkana&Loeb2001}
\begin{align}
V_{c,\mathrm{iso}} = & 23.4 \, \mathrm{km} \, \mathrm{s}^{-1} \, \left( \frac{M}{10^8h^{-1} \, M_\odot} \right)^{2/3} \left( \frac{\Omega_m}{\Omega^z_m}\frac{\Delta_c}{18 \pi^2}\right)^{-1/3}\\ 
& \left(\frac{1+z}{10}\right)^{1/2} \nonumber,
\end{align}
\noindent where \citep{Bryan&Norman1998}
\begin{equation}
\Delta_c = 18\pi^2 + 82d - 39d^2,
\end{equation}
\noindent $d = \Omega^z_m - 1$, and
\begin{equation}
\Omega^z_m = \frac{\Omega_m(1+z)^3}{\Omega_m(1+z)^3+\Omega_\Lambda+\Omega_k(1+z)^2}.
\end{equation}

\subsection{The Galaxy-Halo Connection}

The key galaxy property our model requires is its central velocity dispersion, $\sigma$, which is used in our scaling relation that governs how growth of the SMBH is capped following a major merger.  We calculate $\sigma$ based on the galaxy's stellar mass and effective radius.  Stellar masses are estimated assuming a stellar mass-halo mass relation from abundance matching, while sizes are estimated assuming a stellar mass-size relation. 

\subsubsection{Stellar Masses}

A galaxy's stellar mass is assigned as a function of halo mass and redshift from the abundance matching technique \citep{Moster+2013}.  The mapping is a broken power law that is fit at each epoch.  We note here that our model is particularly sensitive to the slope of the stellar mass-halo mass relation at the high-mass end, which determines the abundance of the most massive and most luminous SMBHs at every epoch.  These models assume an intrinsic scatter of 0.15 dex in these relations.  In our work, this scatter is folded into an overall scattering kernel discussed in \S\ref{ssec:accumulatedScatter}.

\subsubsection{Effective Radii}

At $z=0$, we assume that a galaxy with stellar mass $M_*$ has an effective radius $R_\mathrm{e}$ given by the size-mass relation of \citet{Mosleh+2013}.  Since we are only interested in the size of the galaxy at the end of quasar activity, and we envision a scenario in which quasars quench their hosts, we select the relation followed by red galaxies.  This is given by

\begin{align}
R_\mathrm{e}(M_*,0) = 10^{-0.314} M_*^{0.042} \left(1 + \frac{M_*}{10^{10.537}}\right)^{0.76} \ \mathrm{kpc}.
\end{align}

\noindent For $z > 0$, we assume that this relation evolves at the high mass end \citep{vanDokkum+2010,Huertas-Company+2013}.  We devise a redshift-evolution scheme to match the observations of \citet{Huertas-Company+2013}.  This is parametrised as $R_\mathrm{e}(M_*,z) = R_\mathrm{e}(M_*,0) f(M_*,z)$ with 

\begin{align}
f(M_*,z) = (1+z)^{\gamma(M_*)}
\end{align}

\noindent and

\begin{align}
\gamma(M_*) = \max\left[0, \frac{1}{0.85} \left(\log_{10} (M_*) - 10.75\right)\right] .
\end{align}

\noindent This parametrisation ensures that the mass-size relation does not evolve for low stellar mass hosts with $\log_{10}(M_*) < 10.75$, but $R_\mathrm{e}(z) \propto (1+z)^{-1}$ for $\log_{10}(M_*) = 11.6$, as observed.

Including this redshift evolution increases a high-mass galaxy's velocity dispersion at high redshift, promoting the rapid assembly of high-mass SMBHs.  For the highest-mass galaxies in the model, $\sigma$ is maximised at $z \sim 6$, after which it plateaus.  We comment that increasing the magnitude of this evolution much further can cause the average $\sigma$ of the most massive halos to unphysically shrink with cosmic time. 

\subsubsection{Central Velocity Dispersion}

Given $M_*(M_h,z)$ and $R_e(M_*,z)$ we can now compute $\sigma(M_*,R_e)$.  Our mapping of a galaxy's central velocity dispersion from halo properties is the most important improvement of this work.  

Traditional approaches adopt an isothermal sphere to model the density profile within these dark matter haloes, where the velocity dispersion of an isothermal sphere is given by $\sigma = V_\mathrm{c,iso} / \sqrt{2}$.  This approximation wildly overestimates $\sigma$ for galaxies at the centres of groups and clusters.  We believe this is the primary reason that previous work overestimates the abundance of high-mass black holes at low redshift \citep{Natarajan&Volonteri2012}, and we encourage modelers to take special care when defining $\sigma$ in future work.

\citet{Larkin&McLaughlin2016} develop a framework to calculate $\sigma$ in equilibrium with Jeans modelling assuming radial profiles for stars and the dark matter.  We determine a galaxy's central velocity dispersion by adopting their fitting function given below that takes into account the rearrangement of stars as a function of redshift due to mergers and stellar winds via the parameter $F_{\mathrm{ej}}$.

\begin{align}
\sigma \approx 0.389 \sqrt{\left[ \frac{G M_*}{R_e} \right] \left[ (1 + F_\mathrm{ej}) + \frac{0.86}{f_*(R_e)} \right]}
\end{align} 

\noindent Here, $G$ is the gravitational constant, $M_*$ is the stellar mass, $F_\mathrm{ej}$ is a constant related to the accumulated ejecta from stars over the lifetime of a galaxy, and $f_*$ is the ratio of stars to dark matter within $R_e$.  $M_*$ and $R_e$ are determined as a function of halo mass and redshift as described in the previous two sections.  Then, as in \citet{Larkin&McLaughlin2016}, we set $(1+F_\mathrm{ej}) = 1/0.58$.  Finally, $f_*(R_e)$ requires assuming a profile for both the stars and the dark matter.  We assume that stars follow a \citet{Hernquist1990} profile, appropriate for elliptical galaxies, and the dark matter follows a \citet{Dehnen&McLaughlin2005} profile, with concentration as a function of mass and redshift determined from cosmological dark matter simulations \citep{Dutton&Maccio2014}.  More details of this calculation can be found in \citet{Larkin&McLaughlin2016}.  Although one may not expect the Hernquist profile to be appropriate for the lowest mass galaxies in our model, we find excellent agreement with the observed stellar mass-velocity dispersion relation for quiescent galaxies, confirmed in Appendix \ref{sec:mstar_sigma}.

Figure \ref{fig:Mh_sigma} compares our prescription of a galaxy's central velocity dispersion with that defined for an isothermal sphere at $z=0$.  Clearly an isothermal sphere vastly overestimates $\sigma$ for galaxies at the centres of groups or clusters as noted before.  When mapping from halo mass to SMBH mass, this discrepancy can balloon as $\sigma$ is raised to the fourth or fifth power.  

In Figure \ref{fig:sigmaEvolution}, we trace back the average velocity dispersions as a function of redshift for galaxies that end up in halos of a given mass at $z=0$.  Bootstrapped $1\sigma$ regions from our 15 sets of merger trees are plotted.  Our model is shown on the left, while the evolution of isothermal spheres is shown on the right.  Since the evolution of SMBH mass traces the evolution of $\sigma$, downsizing emerges naturally from our model:  SMBHs in higher mass halos complete their assembly earlier in cosmic time.  The opposite is true in the isothermal sphere model:  $\sigma$ steadily increases with time at the highest masses, while $\sigma$ even unphysically shrinks slightly at the lowest masses.  We note that isothermal sphere models more easily reach high $\sigma$ at $z=6$, but only at the expense of vastly overestimating $\sigma$ at $z=0$.

\begin{figure}
   \centering
   \includegraphics[width=0.45\textwidth]{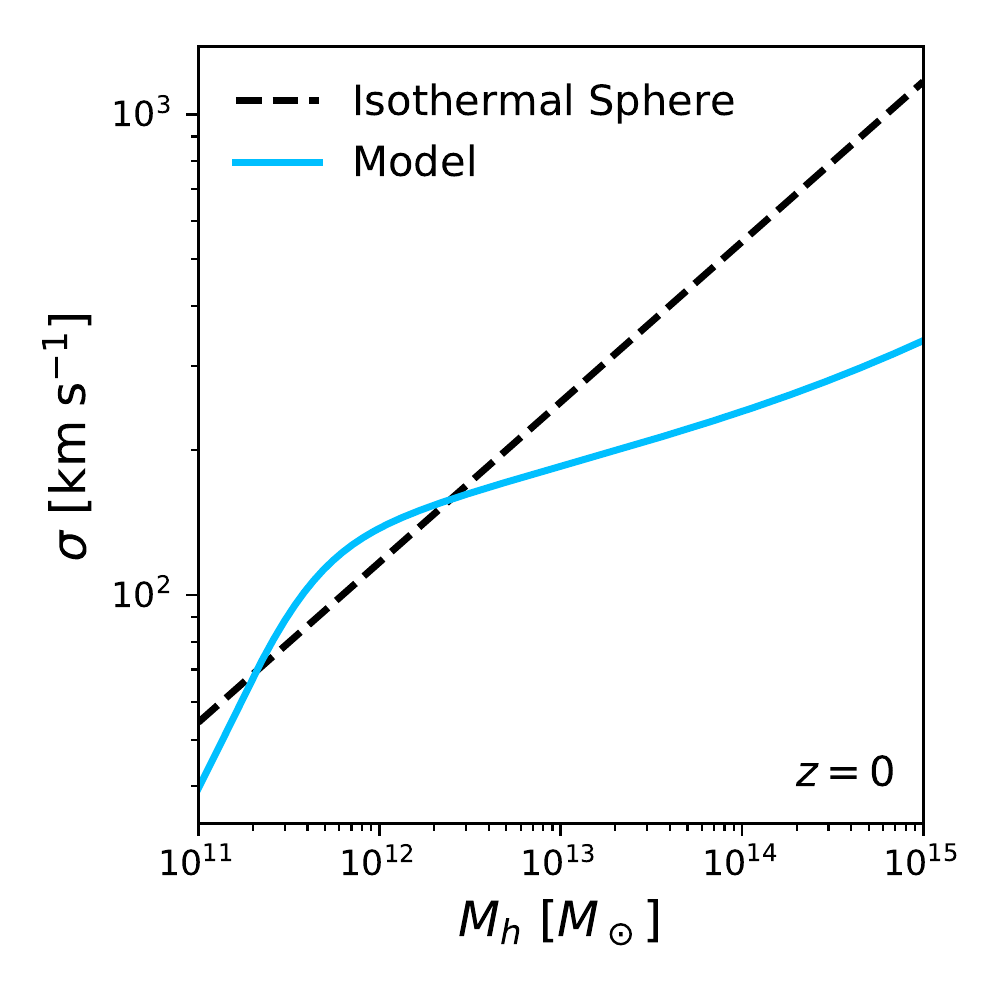}
   \caption{A comparison of central velocity dispersion in our model with more traditional estimates using an isothermal sphere.  The isothermal sphere model vastly overestimates $\sigma$, and therefore $M_\bullet$, in galaxies in groups and clusters.}
   \label{fig:Mh_sigma}
\end{figure}

\begin{figure*}
   \centering
   \includegraphics[width=\textwidth]{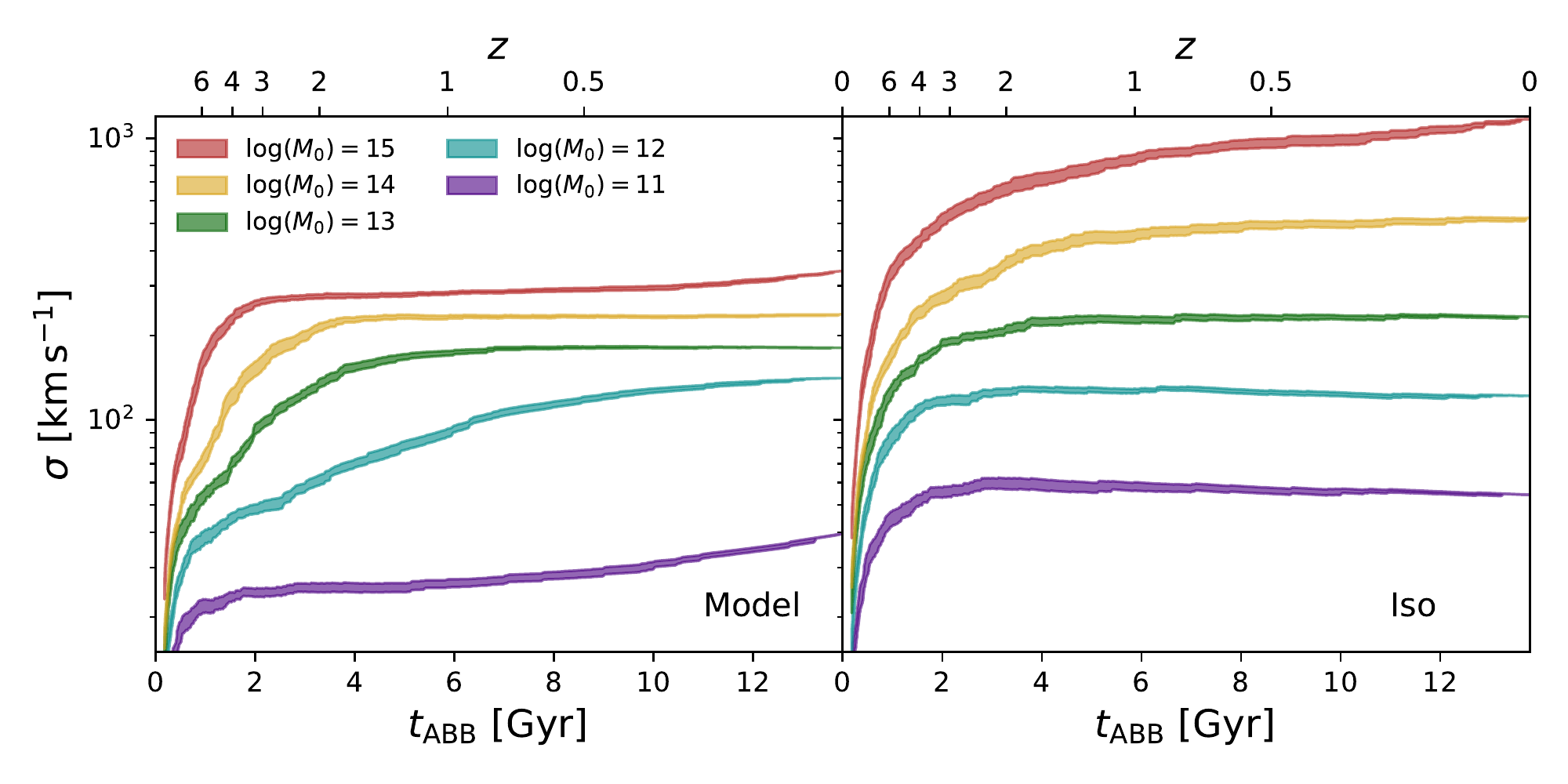}
   \caption{Evolution of velocity dispersion for galaxies that end up in halos of a given mass at $z=0$.  In our model, shown on the left, higher mass halos attain maximum velocity dispersion earlier in cosmic time.  This naturally results in ``downsizing:'' the most massive SMBHs assemble earlier than less massive SMBHs.  The opposite is true of halos modelled as isothermal spheres, shown on the right.}
   \label{fig:sigmaEvolution}
\end{figure*}

\subsection{Dynamics}

Since our picture includes dark matter halo merger triggered accretion as well as black hole mergers , we track the mergers of dark matter halos and relate them to black hole growth.
Dark matter halos merge on the dynamical friction time scale.  At the moment of halo merger, we assume that there is a probability $p_\mathrm{merge}$ that the SMBHs merge as well.  We use a simple implementation of gravitational wave recoil.

\subsubsection{Halo Mergers}

Halo mergers occur on the dynamical friction timescale \citep{Boylan-Kolchin+2008}, given by

\begin{equation}
\frac{t_\mathrm{merge}}{t_\mathrm{dyn}} = A \frac{(M_\mathrm{host}/M_\mathrm{sat})^b}{\ln (1 + M_\mathrm{host}/M_\mathrm{sat})} \exp \left[ c \frac{j}{j_c(E)} \right] \left[\frac{r_c(E)}{r_\mathrm{vir}} \right]^d
\end{equation}

\noindent with constants $A=0.216$, $b=1.3$, $c=1.9$, and $d=1.0$.  Here, $M_\mathrm{host}$ is the mass of the host halo, $M_\mathrm{sat}$ is the mass of the infalling satellite, and $t_\mathrm{dyn}$ is the dynamical time at the virial radius, given by $t_\mathrm{dyn} =  0.1 H(z)^{-1}$.  For all orbits, we assume typical values of the circularity and orbital energy, given by $j/j_c(E) = 0.5$ and $r_c(E)/r_\mathrm{vir} = 0.6$.  

A merger between two halos is considered ``major'' if their halo mass ratio is at least 1:10, and ``minor'' otherwise. In our scheme, if a minor merger occurs the satellite is not tracked any further.  That is, it falls out of our book-keeping.  It is assumed that such minor mergers do not trigger accretion for the central SMBH, and that the SMBH of the smaller galaxy is left wandering in the outskirts of the halo, never to accrete again. 

\subsubsection{Black Hole Mergers}
\label{ssec:bh_mergers}

Once a major merger occurs, we assume that there is a probability $p_\mathrm{merge}$ that the SMBHs also merge, at exactly the same time.  We find that the parameter $p_\mathrm{merge}$ is required to suppress the amount of growth due to SMBH mergers at low redshift.  Such a parameter makes physical sense, as there will be delays in black hole mergers post host halo mergers due to intra-galaxy dynamics. In our model, high-mass SMBHs finish assembling early in cosmic time, and SMBH mergers can dominate their growth at low redshift without the addition of this parameter.  We tune $p_\mathrm{merge} = 0.1$ in order to allow for the rapid assembly of the most massive SMBHs at $z \sim 6$ without overgrowing them by $z=0$.

Determining the time delay between halo merger and black hole merger is a subject of active research.  On galactic scales, the time between galaxy merger and binary formation requires accurate treatment of dynamical friction in a cosmological context.  Recently, \citet{Tremmel+2017b} have computed these time intervals in the Romulus simulations \citep{Tremmel+2017a}, where the dynamical friction on black holes is carefully determined.  Including a more robust treatment of SMBH dynamics will be the subject of future work.

\subsubsection{Gravitational Wave Recoil}

In the early universe, gravitational wave recoil is capable of ejecting black holes from their host halos \citep{Yoo&Miralda-Escude2004}.  When two SMBHs merge in our model, we compute the gravitational wave kick speed according to a fitting formula derived from general relativistic black hole merger simulations \citep{Lousto+2012}.  This formula requires the spins of each SMBH, as well as the angles between each spin vector and the angular momentum of the orbit.  In our prescription, these angles are each drawn randomly from a uniform distribution between 0 and $2\pi$.  We also do not track the spin evolution of SMBHs due to accretion, as doing so requires ad-hoc assumptions about the alignment of the accretion disk with the SMBH spin axis.  Hence, when a SMBH is seeded, its spin parameter is also drawn from a uniform distribution between 0 and 1, and only future SMBH mergers are allowed to alter its spin.

Once the ejection velocity is calculated, it is compared to the escape velocity of the galaxy, assumed to be $v_\mathrm{esc} = 2 V_c(M_h,z)$.  If the ejection velocity exceeds the host halo's escape velocity, the SMBH is ejected from the galaxy, and it joins a ``wandering'' population of SMBHs.  We do not track the orbital decay of these wandering SMBHs.

In our model, recoil does not have a strong effect on SMBH assembly for two reasons.  First, a SMBH's growth in our model tends to be limited by halo properties rather than its own mass.  Hence, if a SMBH is ejected, the next one to take its place will grow to the same final mass (see \S\ref{sssec:fuelling_burst}).  Second, the number of mergers in our model have been suppressed due to the adoption of the parameter $p_{\rm merge}$ to take into account the dynamical delays expected between halo mergers, galaxy mergers and black hole mergers (see \S\ref{ssec:bh_mergers}). 

\subsection{Black Hole Seeding}

There are currently two broad categories of SMBH seeding models.  Light seeds are hypothesised to originate from the remnants of Pop III stars.  These are naturally generated from stellar evolution, but the masses of Pop III stars  and therefore their remnants have been decreasing in the latest simulations \citep{Hirano+2014}, making them difficult to grow to $10^9 \ M_\odot$ by $z=6-7$ \citep{Alvarez+2009}.  Heavy seeds are thought to originate from the direct collapse of primordial gas clouds.  Such DCBHs require special physical conditions, such as metal-free gas and Lyman-Werner photons that dissociate H$_2$ molecules \citep[e.g.,][]{Agarwal+2016}, but can be much more massive than seeds generated by Pop III stars.   

Here, we focus only massive seeds -- primarily DCBHs or perhaps the Pop III remnants that have extremely rapid early growth via processes suggested by \citet{Alexander&Natarajan2014}.  Seeds are allowed to form in our model within the redshift range $z \in [15,20]$ in halos which lack progenitors.  We use the DCBH model of \citet{Lodato&Natarajan2006,Lodato&Natarajan2007} to determine seed masses.  In this model, mass infall to the centre of a dark matter halo is driven by non-axisymmetric spiral structures in the proto-galactic gas disk.  The central mass accumulation, assumed in our model to be the black hole mass, is given by
\begin{equation}
M_\bullet = m_dM_h \left[1 - \sqrt{\frac{8 \lambda}{m_d Q_c} \left(\frac{j_d}{m_d}\right) \left(\frac{T_\mathrm{gas}}{T_\mathrm{vir}}\right)^{1/2}} \right]
\end{equation}
\noindent if
\begin{equation}
\lambda < \lambda_\mathrm{max} \equiv m_d Q_c / 8 (m_d/j_d)(T_\mathrm{vir}/T_\mathrm{gas})^{1/2}
\end{equation}
\noindent for the disk to be gravitationally unstable, and
\begin{equation}
T_\mathrm{vir} < T_\mathrm{max} \equiv T_\mathrm{gas} \left(\frac{4\alpha_c}{m_d}\frac{1}{1+M_\bullet/m_dM_h}\right)^{2/3}
\end{equation}
\noindent to prevent fragmentation.  Here, $m_d$ is the fraction of total halo mass in the disk, $M_h$ is the mass of the halo, $\lambda$ is the halo spin parameter, $j_d$ is the fraction of the halo angular momentum of the disk, $T_\mathrm{gas}$ is the gas temperature, $T_\mathrm{vir}$ is the halo virial temperature, $\alpha_c$ is a dimensionless parameter representing the critical gravitational torque above which a disc fragments, and $Q_c$ is the critical Toomre parameter above which the disk is assumed to be stable.  Based on the results of previous studies, we set $m_d = j_d = 0.05$ \citep{Mo+1998}, $\alpha_c = 0.06$ \citep{Rice+2005}, and $Q_c = 3$\footnote{The value of the Q-parameter that essentially tunes the formation efficiency of DCBHs in their work was taken to be 2. Here we have increased it to $Q=3$ in order to increase the occupation fraction of SMBHs at the low-mass end.} \citep{Volonteri+2008}.  $T_\mathrm{gas}$ is assumed to be $5000 \ K$, appropriate for gas without metal coolants.  The first of these conditions ensures that the disk has low enough angular momentum to trigger infall, while the second of these conditions ensures that the disk does not fragment.  If either of these conditions is not satisfied, a black hole is not seeded.

Refinements have been developed to our understanding of seeding that are absent in our model due to the lack of spatial information in our merger trees.  In particular, the formation of a DCBH requires a nearby Lyman-Werner source that suppresses H$_2$ formation \citep{Oh&Haiman2002}. The suppression of the formation of molecular hydrogen prevents cooling and fragmentation as required. The halo must also be free of metal pollutants to prevent fragmentation \citep{Clark+2008,Ferrara+2014}.  Finally, it was recently noticed that the tidal field induced by the source of the radiation field may disrupt a halo that would otherwise be seeded \citep{Chon+2016}.

\subsection{Accretion and Radiation}

A SMBH's accretion rate is parameterized in terms of its Eddington ratio, $f_\mathrm{Edd}$, defined by $f_\mathrm{Edd} \equiv \dot{M}/\dot{M}_\mathrm{Edd}$, where
\begin{equation}
\dot{M}_\mathrm{Edd} = \frac{4 \pi G M_\bullet m_p}{\epsilon \sigma_T c} \equiv \frac{M_\bullet}{\epsilon t_\mathrm{Edd}}.
\end{equation}
\noindent Here, $G$ is the gravitational constant, $m_p$ is the proton mass, $\sigma_T$ is the Thompson cross-section, $c$ is the speed of light, and $\epsilon$ is the radiative efficiency, which we set to a constant value of 0.1.  Combining constants, $t_\mathrm{Edd} \equiv 450$ Myr.  Consequently, if a SMBH accretes at an Eddington ratio of $f_\mathrm{Edd}$ from time $t$ to $t + \Delta t$, its mass changes via
\begin{equation}
M(t+\Delta t) = M(t) \exp\left(f_\mathrm{Edd}\frac{1-\epsilon}{\epsilon} \frac{\Delta t}{t_\mathrm{Edd}}\right)
\end{equation}
\noindent while radiating with a bolometric luminosity of
\begin{equation}
L = \epsilon f_\mathrm{Edd} \dot{M_\mathrm{Edd}}c^2.
\end{equation}

\subsection{Fuelling}
\label{ssec:fuelling}

We explore several options for our fuelling prescriptions.  We implement two fuelling modes, a burst mode triggered by major mergers of the dark matter haloes, and a steady mode that operates otherwise.

\subsubsection{Burst Mode}
\label{sssec:fuelling_burst}

Each SMBH is initialised with an accretion budget of $0 \ M_\odot$.  When a major merger occurs, an amount of mass equal to $M_\mathrm{cap}(\sigma) - M_\bullet$ is added to the SMBHs accretion budget.  We define 

\begin{align}
\log (M_\mathrm{cap}(\sigma)) \equiv \alpha + \beta \left( \frac{\sigma}{200 \; \mathrm{km}\;\mathrm{s}^{-1}} \right) \label{eqn:mass_cap},
\end{align}

\noindent and tune the parameters $\alpha$ and $\beta$ in order to reproduce the local $M_\bullet-\sigma$ relation, setting $\alpha=8.45$ and $\beta=5$.  We note that $\beta=5$ is appropriate for energy-driven wind feedback \citep{Haehnelt+1998,Natarajan&Treister2009,King2010}.

While a SMBH's accretion budget is non-zero, it accretes this mass at an Eddington ratio given by 

\begin{align}
f_\mathrm{Edd} = \mathrm{min} \left[1, \ln \left(\frac{M_\mathrm{budget}}{M_\bullet} + 1\right) \frac{t_\mathrm{Edd}}{\Delta t}\frac{\epsilon}{1-\epsilon} \right].
\end{align}

That is, it depletes the budget as quickly as possible during a time step, up to the Eddington limit.  Our model does not require super-Eddington episodes at high redshift, since we assume that accretion can continue uninterrupted in our models until $M_\mathrm{cap}$ is reached.  At the highest masses, it is $M_\mathrm{cap}$, rather than the Eddington limit, that limits SMBH growth. 

\subsubsection{Steady Mode}

There are two reasons to include a second mode of growth in the model.  First, the burst mode drastically underestimates the AGN luminosity function at redshifts $z \leq 1$.  Second, the burst mode fails to grow SMBHs in halos of mass $M_h < 10^{12} \ M_\odot$ to their appropriate mass.  We explore two simple parameterisations for this additional mode to address each of these issues separately.

\begin{itemize}
\item A Universal Power Law:  It has been shown that an Eddington ratio distribution independent of stellar mass is successful at reproducing the X-ray luminosity function at redshifts $z \lesssim 1.2$ \citep{Jones+2017}.  When this steady mode is enabled, a SMBH with a depleted accretion budget is assigned a random Eddington ratio drawn from a power law distribution.  We assume that this power law is universal, independent of both $M_*$ and $z$.  We tune this power law by-eye to roughly match the $z=0.1$ bin of the bolometric luminosity functions.  It is parametrised as follows

\begin{align}
\frac{dp}{d \log f_\mathrm{Edd}} \propto f_\mathrm{Edd}^{\gamma}
\end{align}

\noindent for $\log f_\mathrm{Edd} \in [-4, 0]$ and $\gamma = -0.9$.  

This parameterisation, though it fails to properly describe the redshift-evolution of the AGN luminosity function, provides insight into a steady mode which does not dominate SMBH growth.  Readers interested in estimates of redshift-evolving Eddington ratio distributions, will find more robust treatments in population synthesis models \citep{Merloni&Heinz2008}.  Redshift-evolving distributions of specific accretion rates have also been derived from multi-wavelength surveys \citep[e.g.,][]{Aird+2017}, but these ``instantaneous'' accretion rates may differ from the longer term accretion rates relevant for SMBH growth due to AGN stochasticity \citep{Hickox+2014}. 

\item The AGN Main Sequence:  Stacked observations of star-forming galaxies reveal evidence for the existence of an ``AGN Main Sequence,'' with SMBH accretion proportional to star formation rate (SFR) \citep{Mullaney+2012}.  In this model, we estimate the SFR of a SMBH host galaxy by comparing its change in stellar mass between two time steps and subtracting any mass gained via mergers.  At every time step, once a 
SMBH has consumed the gas in its allocated budget $(M_{\rm cap} -M_{\bullet})$ it is still allowed to continue accreting as part of implementing the long-lived steady mode, at a rate equal to $\mathrm{SFR} / 10^{3}$ until the next merger.
\end{itemize}

\subsection{Modelling Accumulated Scatter}
\label{ssec:accumulatedScatter}

In this model, we have bridged the gap between halo mass and black hole mass through the use of several scaling relations.  To summarise, we use the stellar mass-halo mass relation to determine $M_*(M_h,z)$, the size-mass relation to determine $R_e(M_*,z)$, and the $M_\bullet-\sigma$ relation to determine $M_\mathrm{cap}(\sigma)$.  Thus far, we have neglected the scatter intrinsic to these relations.  However, properly accounting for the scatter accumulated through these maps is crucial to accurately calculating mass and luminosity functions.

We model scatter when determining mass and luminosity functions by convolving our raw histograms with a scattering kernel.  This kernel is parameterised as a log-normal with a fixed width $\sigma_s$.  To determine the appropriate value for $\sigma_s$, we calibrate our model to the local SMBH mass function.  Following our scaling relations, we calculate the local mass function analytically.  The amount of scatter is then tuned by-eye to obtain a reasonable match to the range of acceptable mass function estimates \citep{Shankar+2009}.  We then arrive at $\sigma_s = 0.3$ dex.  Additional details of this calculation can be found in Appendix \ref{sec:scatter}.

\subsection{Four Model Variants}

\begin{table*}
\centering
\begin{tabular}{lcccccc}
\hline
& \multicolumn{2}{c}{Burst} & \multicolumn{2}{c}{Steady:  Power Law}     & Steady:  AGN Main Sequence & Dynamics    \\
\hline
Model & $\alpha$       & $\beta$       & $\log f_\mathrm{Edd,min}$ & $\gamma$ & $\dot{M}_\bullet / \mathrm{SFR}$ & $p_\mathrm{merge}$ \\
\hline
Burst & 8.45           & 5             &                         &                 &  & 0.1 \\          
PowerLaw & 8.45           & 5             & -4                        & -0.9                &  & 0.1 \\          
AGNMS & 8.45           & 5             &                      &                 & $10^{-3}$ & 0.1 \\          
Iso & 8.00           & 4             & -4                        & -0.9                &  & 0.1 \\          
\hline
\end{tabular}
\caption{Parameters tuned in our model for the four model variants discussed in this work.  Burst parameters are tuned to $M_\bullet-\sigma$, power law parameters are tuned to the local luminosity functions, the AGN main sequence parameter is tuned to roughly match stacked observations, and the dynamics parameter is tuned to match $M_\bullet-\sigma$ at high masses.}
\label{tab:parameters}
\end{table*}

In this paper, we compare and contrast 4 different models, described below.

\begin{itemize}
\item Burst:  This model uses the burst mode without an accompanying steady mode.
\item PowerLaw:  This model uses the burst mode together with the universal power law steady mode.
\item AGNMS:  This model uses the burst mode together with the AGN main sequence steady mode.
\item Iso:  This model uses the same accretion prescriptions as PowerLaw, but with the central velocity dispersion set to that of an isothermal sphere, $\sigma = V_c/\sqrt{2}$.
\end{itemize}

These models each use 3-5 tuned parameters.  We summarise our choices in Table \ref{tab:parameters}.  To recap, $\alpha$ and $\beta$ are tuned to reproduce the local $M_\bullet-\sigma$ relation.  The power law parameters are tuned to reproduce the $z=0.1$ luminosity function, while the AGN main sequence parameter roughly matches stacked observations and grows low-mass SMBHs.  Finally, $p_\mathrm{merge}$ suppresses SMBH mergers so that our various models can reproduce the local $M_\bullet-\sigma$ relation at the high-mass end.

\begin{figure*}
   \centering
   \includegraphics[width=\textwidth]{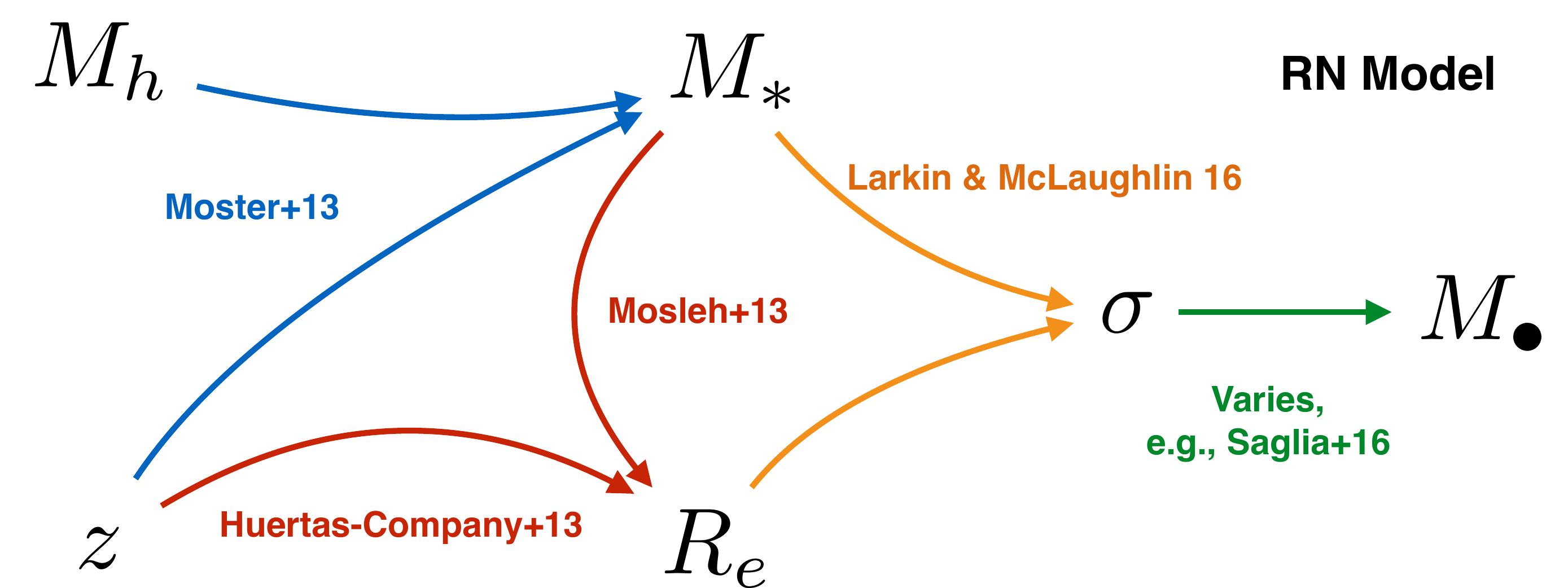}
   \includegraphics[width=\textwidth]{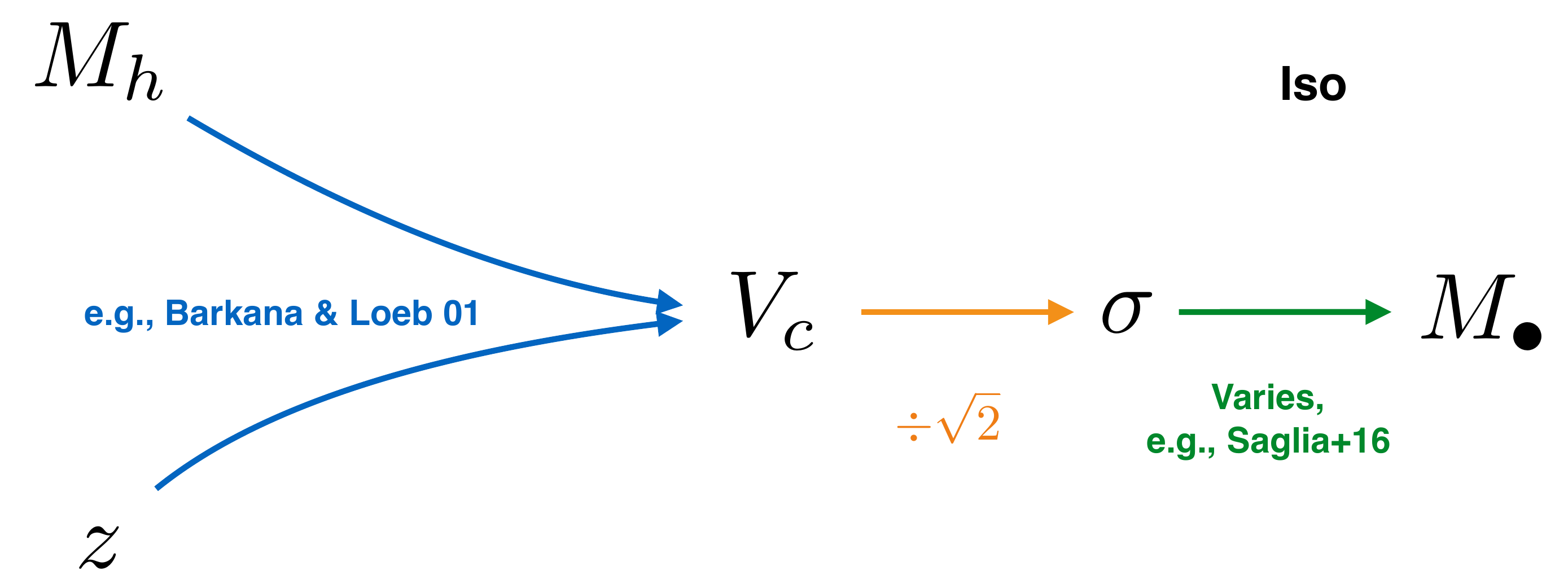}
   \caption{From halo properties to black hole mass.  This figure sketches the modelling steps taken to map from halo mass and redshift to (maximum) black hole mass.  In our model, we compute central velocity dispersion from inferred galaxy properties.  In the Iso model, and in many previous works, the central velocity dispersion is directly proportional to the circular velocity from the spherical collapse model.}
   \label{fig:schematics}
\end{figure*}

Figure \ref{fig:schematics} schematically summarises the modelling steps taken to map from halo mass and redshift to the maximum black hole mass for the burst mode in our model, as described in the previous sections.  The Iso model, and many previous works, define $\sigma$ based on the spherical collapse model, whereas we attempt to define $\sigma$ based on inferred galaxy properties.

\section{Results}
\label{sec:results}

Our aim is to match the $M_\bullet-\sigma$ relation, the local mass function, redshift-evolving luminosity functions, and the mass function of BLQs.  Simultaneously matching all of these observational data is the key challenge.  In practice, the local $M_\bullet-\sigma$ relation is easiest to match, requiring the tuning of only $\alpha$ and $\beta$ of the mass cap (at least at the high-mass end).  If the $M_\bullet-\sigma$ relation is matched, then the local mass function for SMBHs is matched if and only if the mapping $M_h \mapsto \sigma$ is correct.  Luminosity functions are more difficult to match, since this requires SMBHs to not only accumulate mass and grow but also radiate self-consistently.  Mass functions for BLQs are the final step in difficulty, since the probability of a SMBH being observed as a BLQ is a function of both its luminosity and its Eddington ratio.  We find that none of the models discussed here can match all of these observables perfectly.  Of course, if one chooses to, parameters can be fine-tuned and added to provide an excellent match to a given observable, but this is not the purpose of this exercise.  Given our limited set of physically motivated parameters, we discuss the challenges in matching each component, and determine which parameters drive the fit.

\subsection{The $M_\bullet-\sigma$ Relation}
\label{ssec:msigma}

\begin{figure*}
   \centering
   \includegraphics[width=0.8\textwidth]{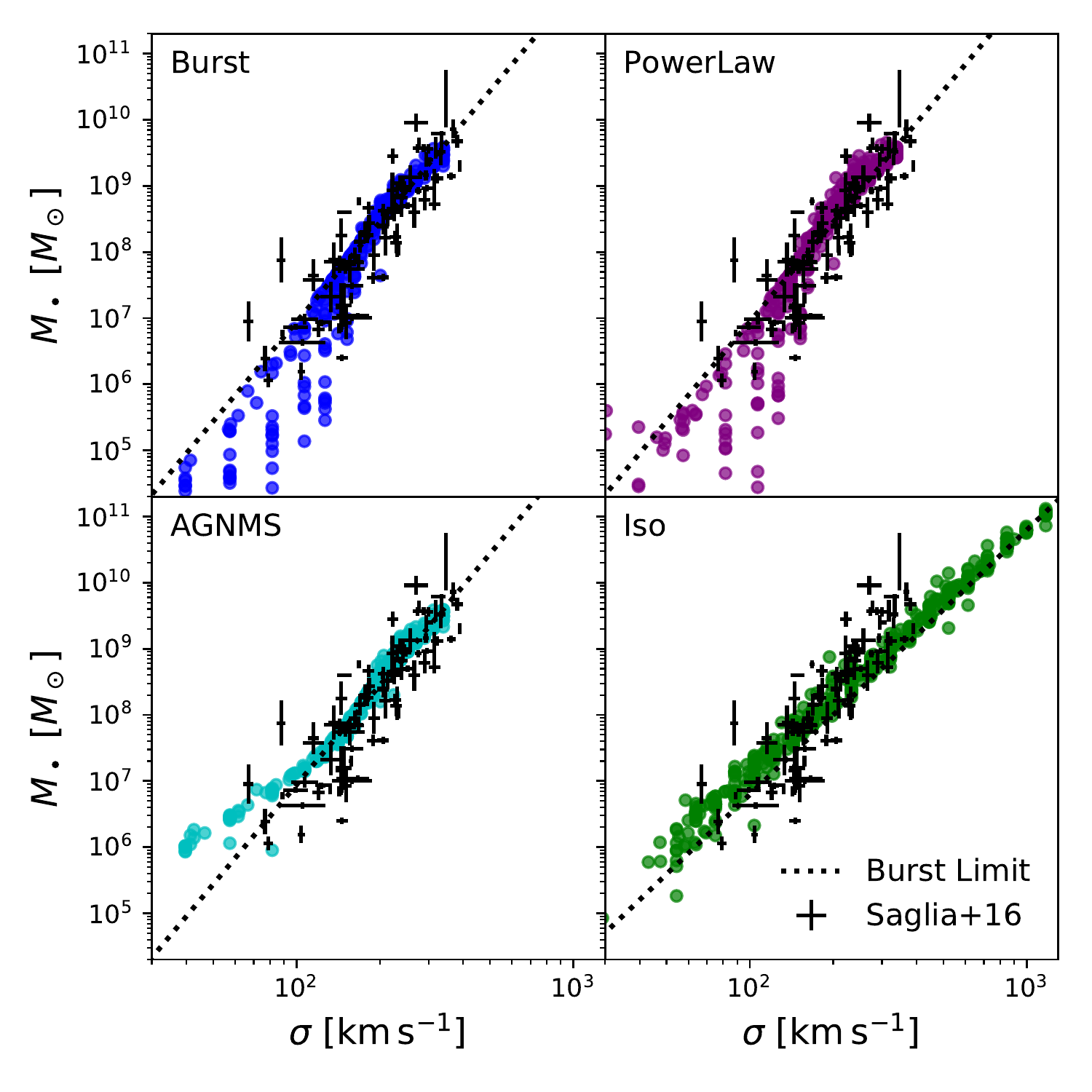}
   \caption{The local $M_\bullet-\sigma$ relation in our SAM compared to observational data for our four models.  Burst, PowerLaw, and AGNMS agree at the high mass end, but only AGNMS can consistently grow SMBHs at the low-mass end to the $M_\bullet-\sigma$ relation.  The Iso variant extends to far higher $\sigma$, and therefore $M_\bullet$, since this variant overestimates velocity dispersion in groups and clusters.}
   \label{fig:msigma}
\end{figure*}

In Figure \ref{fig:msigma}, we compare the $M_\bullet-\sigma$ relation obtained from our SAMs with a recent compilation of dynamically measured SMBH masses by \citep{Saglia+2016}.   The SMBHs plotted here include all non-wandering SMBHs at $z=0$, which contains those in ongoing halo mergers.  First, let us consider the Burst variant, which includes our improved model for $\sigma(M_h,z)$. This model is able to reproduce the massive end of the $M_\bullet-\sigma$ relation, which indicates that merger-driven, feedback-limited bursts are sufficient to explain the assembly of these SMBHs.  

However, below a mass of approximately $M_h \sim 10^{12} \ M_\odot$, there is considerable scatter:  some SMBHs make it to the $M_\bullet-\sigma$ relation, but most fall short.  This is due to the low-mass evolution of $\sigma(z)$ as shown in Figure \ref{fig:sigmaEvolution}.  For low-mass halos, most of the evolution in $\sigma$ occurs at low-redshift, when mergers occurs less frequently.  Since our model requires a major merger to kick-start the burst mode, SMBHs in halos that experience quieter merger histories at low-redshift are more offset from the $M_\bullet-\sigma$ relation.  

Therefore, a second fuelling channel---the steady accretion mode---is required to make up the difference for low-mass SMBHs.  The PowerLaw variant has a steady Eddington ratio distribution tuned to match the $z=0.1$ luminosity function, as will be discussed in \S\ref{ssec:luminosityFunctions}.  Yet this variant does no better than Burst at growing low-mass SMBHs.  That is, the same universal power law that can be used to match the local luminosity function cannot also be responsible for growth at the low-mass end.  The AGNMS variant does a much better job at assembling low-mass SMBHs, albeit with a flattening of the slope at low masses.  This slope change is a reflection of the fact that $10^{-3} M_* > M_\mathrm{cap}(\sigma)$ for low masses.  However, we shall see in \S\ref{ssec:luminosityFunctions} that although this growth mode is successful at matching the $M_\bullet-\sigma$ relation, it fails to reproduce bolometric luminosity functions.

In the Iso model, we have tuned $\alpha$ and $\beta$ to values which keep the local SMBH mass density within reason.  The compromise is necessitated by the fact that Iso predicts much larger values of $\sigma$, and therefore $M_\bullet$, for group and cluster scales.  Interestingly, Iso does not have the same problem at low masses that our improved model does, due to the fact that $\sigma(z)$ actually peaks early in the universe (and then unphysically shrinks) for the typical $10^{11} \ M_\odot$ halo (see Figure \ref{fig:sigmaEvolution}).

\subsection{Mass Functions}
\label{ssec:massFunctions}

\begin{figure}
   \centering
   \includegraphics[width=0.5\textwidth]{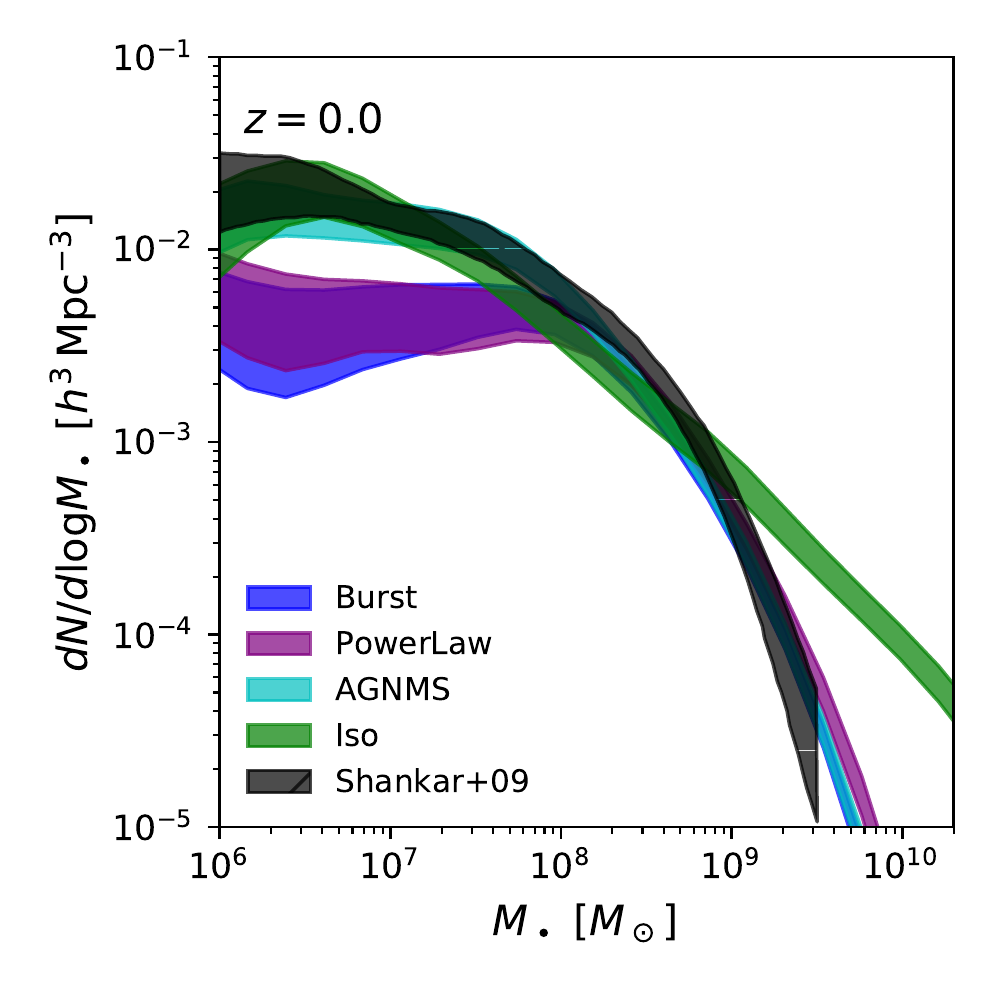} 
   \caption{Local mass functions generated by our SAM compared to a compilation of estimates using various methods \citep{Shankar+2009}.  Burst and PowerLaw underproduce low-mass SMBHs.  AGNMS best matches the $M_\bullet-\sigma$ relation, and therefore produces the most accurate mass function.  Iso overproduces high-mass SMBHs.}
   \label{fig:mass_local}
\end{figure}

\begin{figure*}
   \centering
   \includegraphics[width=\textwidth]{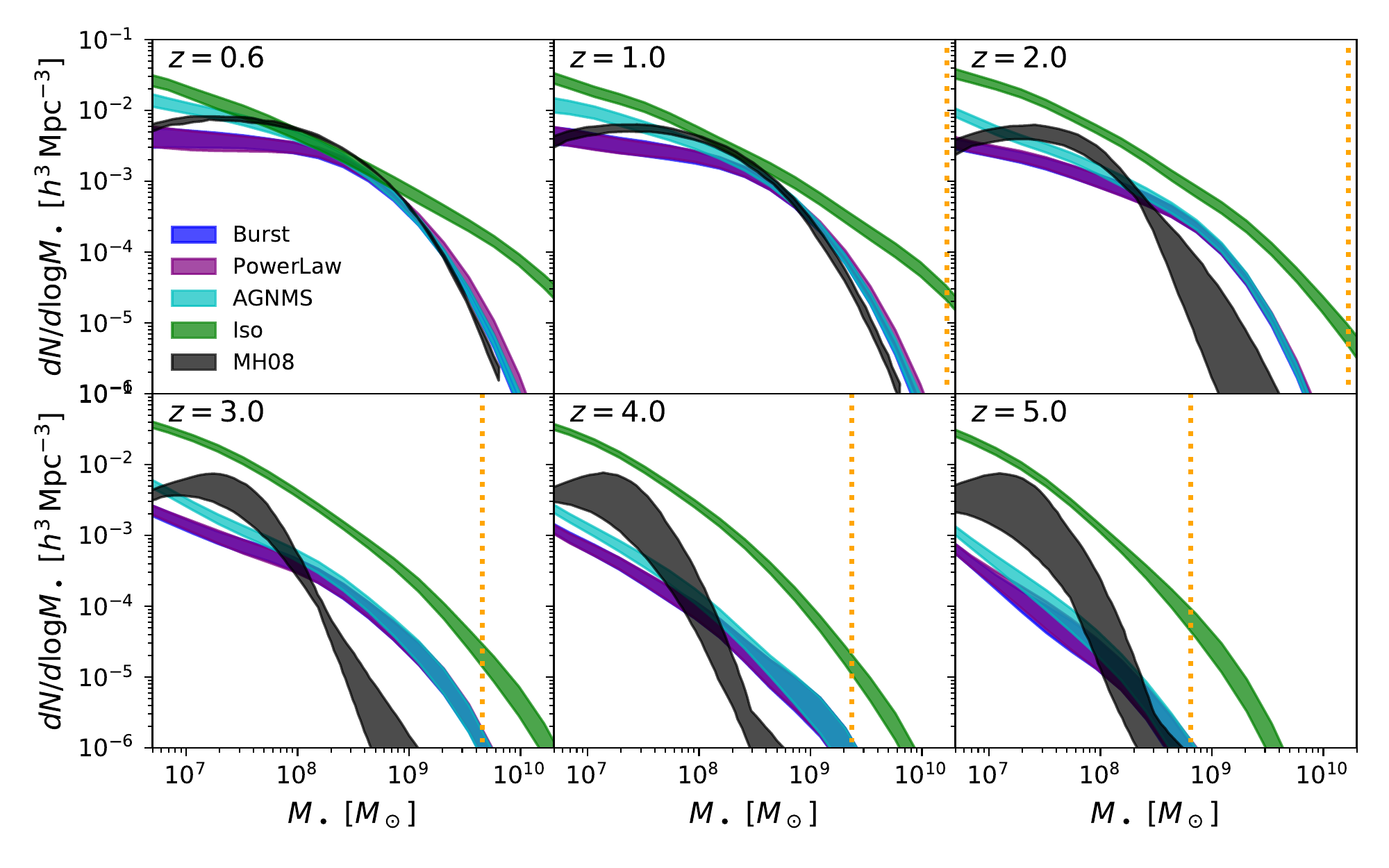} 
   \caption{Mass functions generated by our SAM compared to the population synthesis model of \citet{Merloni&Heinz2008}.  Burst, PowerLaw, and AGNMS produce high masses earlier and low masses later.  Iso assembles most SMBHs early in the universe, and grows the high-mass end last.  Note that Burst and PowerLaw almost completely overlap.}
   \label{fig:mass}
\end{figure*}

SMBH mass functions provide a view of the connection between SMBHs and their hosts, since each SMBH is weighted by the abundance of its host halo. Once the $M_\bullet-\sigma$ relation is reproduced and the mapping from $M_h$ to $\sigma$ is correct, then the local mass function is matched.  We compute SMBH mass functions in our SAM as a function of redshift by binning SMBHs weighted by the number density of their $z=0$ host halos.  Since we generate 15 trees each of the same halo mass, these are used to bootstrap error ranges. With this done, the mass function is convolved with our scattering kernel, a lognormal of width 0.3 dex, in order to account for the accumulated scatter from the scaling relations we use (see \S\ref{ssec:accumulatedScatter}).  

First, we display the local mass functions inferred from our model in Figure \ref{fig:mass_local}.  We compare our results to a compilation of recent estimates of the local SMBH mass function, since the choice of scaling relations impacts estimates of the local black hole mass function \citep{Shankar+2009}.  At the high-mass end, all models except Iso are in excellent agreement.  Yet below $M_\bullet \sim 10^{8} \ M_\odot$, both our Burst and PowerLaw models fall short.  This is a reflection of their inconsistency in growing low-mass SMBHs discussed in \S\ref{ssec:msigma}.  The AGNMS variant, the one which best matches $M_\bullet-\sigma$, produces the most accurate mass function.

Next, we compare the redshift-evolution of our mass functions to the population synthesis model of \citet{Merloni&Heinz2008}, shown in Figure \ref{fig:mass}.  \citet{Merloni&Heinz2008} work backwards, integrating the local mass function to higher redshift with bolometric luminosity functions.  Note that uncertainties in estimating the local mass function are not taken into account in \citet{Merloni&Heinz2008}, and error regions reflect only the uncertainty inherent in their modelling.  Our mass and luminosity functions include estimates for the completeness of our merger trees.  Shown here as a dotted orange line, we display the SMBH mass above which our trees are missing at least 50 per cent of SMBH hosts for our non-Iso models.  More details about this completeness calculation can be found in Appendix \ref{sec:completeness}. 

As redshift increases, all model variants predict different evolution from the population synthesis model of Merloni \& Heinz.  The Burst, PowerLaw, and AGNMS models evolve very similarly.  These variants produce high mass SMBHs earlier and low-mass SMBHs later compared to the population synthesis model.  This is likely an improvement, since at high redshift, this population synthesis model predicts fewer high-mass SMBHs than we have observed BLQs \citep{Natarajan&Volonteri2012}.  The Iso model predicts a much larger abundance of SMBHs of all masses, especially at the high-mass end.  This is because SMBH growth tends to occur earlier in cosmic time in this variant (see Figure \ref{fig:sigmaEvolution}), in addition to the fact that this model produces many SMBHs with erroneously high masses. 

\subsection{Bolometric Luminosity Functions}
\label{ssec:luminosityFunctions}

\begin{figure*}
   \centering
   \includegraphics[width=\textwidth]{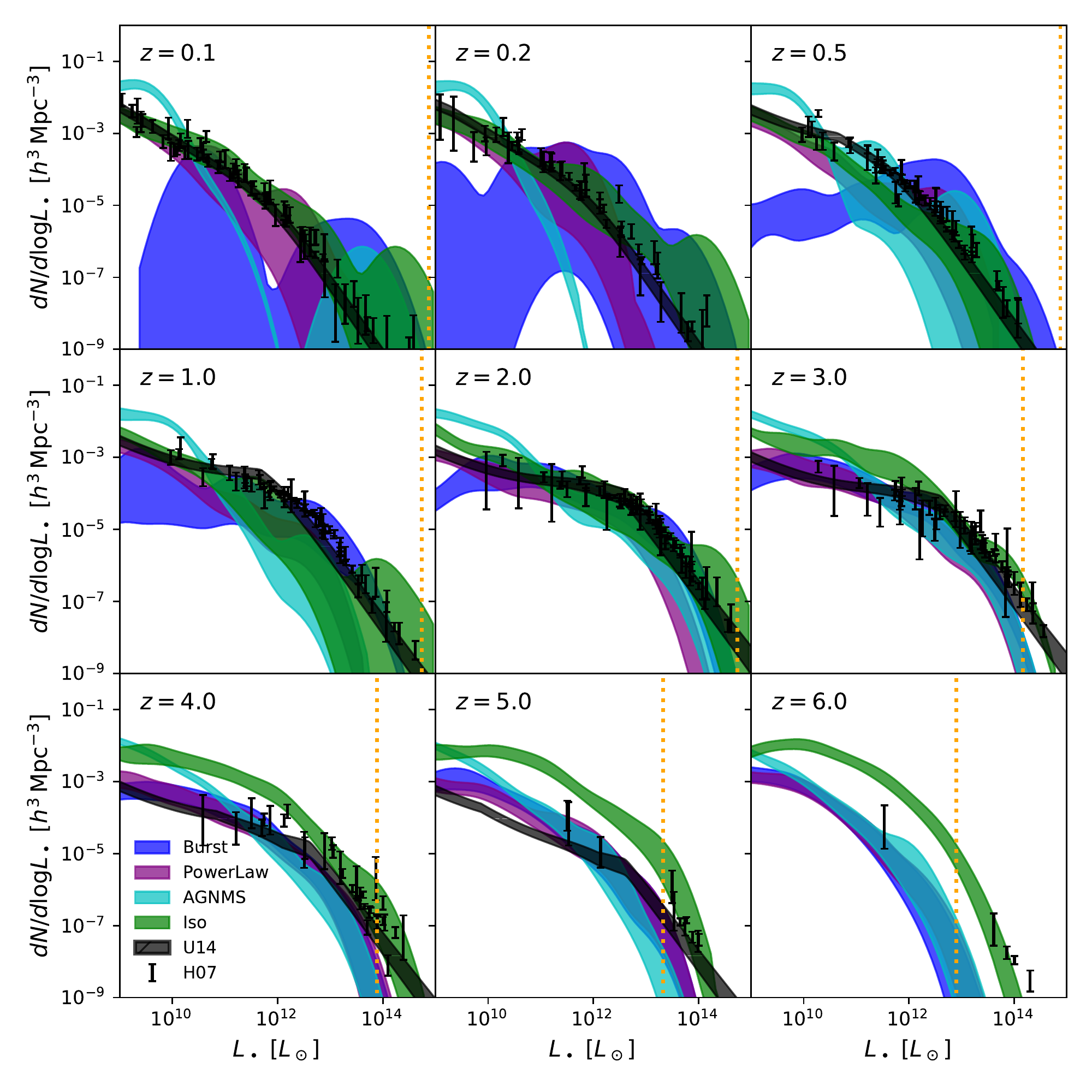} 
   \caption{Bolometric luminosity functions derived from our SAM compared to estimates in the literature \citep{Hopkins+2007,Ueda+2014}.  PowerLaw and AGNMS broadly agree with Burst for $z \geq 2$, indicating that merger-triggered, feedback-limited growth is sufficient to explain the assembly of SMBHs until the peak of AGN activity.  At lower $z$, the steady mode dominates and these three models diverge.  PowerLaw, tuned to the $z=0.1$ slice, agrees best overall, whereas AGNMS does not produce low-z, high-L AGN.  The Iso model fares surprisingly well for $z \leq 2$, even though it overestimates the abundance of high-mass SMBHs.  For higher $z$, Iso lies above the observations, indicating that it assembles SMBHs too early in cosmic time.}
   \label{fig:lum}
\end{figure*}

Using similar techniques, we also generate bolometric luminosity functions, shown in Figure \ref{fig:lum}.  We compare to estimates of the bolometric luminosity function by \citet{Hopkins+2007} and \citet{Ueda+2014}.  Here, our completeness estimates are equal to the Eddington luminosity of our limiting SMBH mass.  

First, the Burst model is surprisingly capable of matching the luminosity functions for $z \geq 2$.  This indicates that only the burst mode, with appropriate mass limits, is important for generating luminosity functions until the peak of AGN activity.  However, as $z$ decreases further, the burst mode becomes increasingly insufficient for triggering AGN.  This motivates the existence of a second mode of growth that begins to dominate at low-redshifts. 

The PowerLaw model is tuned to provide a reasonable match at $z=0.1$.  Among our four models, it provides the best match to these luminosity functions overall.  It is most discrepant at the high-luminosity end for redshifts between $z=0.1$ and $z=2$.  This indicates that a universal power-law independent of both $M_\bullet$ and $z$ cannot reproduce the luminosity function at all redshifts, which is unsurprising.  Indeed, the Eddington ratio distribution is thought to depend both on stellar mass and redshift \citep{Aird+2017}.  The AGNMS variant most faithfully reproduces the $M_\bullet-\sigma$ relation, yet this mode is clearly unable to reproduce the luminosity functions for $z<2$.  In particular, it fails to produce low-redshift luminous quasars, since the hosts of the most massive SMBHs are not star-forming.  AGNMS also overproduces the low-luminosity end at all redshift, a reflection of the flattening in the $M_\bullet-\sigma$ relation at the low-mass end seen in \S\ref{ssec:msigma}.

Finally, using the same steady mode as PowerLaw, Iso performs surprisingly well for $z \leq 2$.  Given that the mass function is incorrect, however, we consider this a coincidence.  For $z \geq 3$, Iso lies above the observations, mirroring the behaviour of their mass functions.

\subsection{Broad Line Quasar Mass Functions}
\label{ssec:massFunctions_blq}

\begin{figure*}
   \centering
   \includegraphics[width=\textwidth]{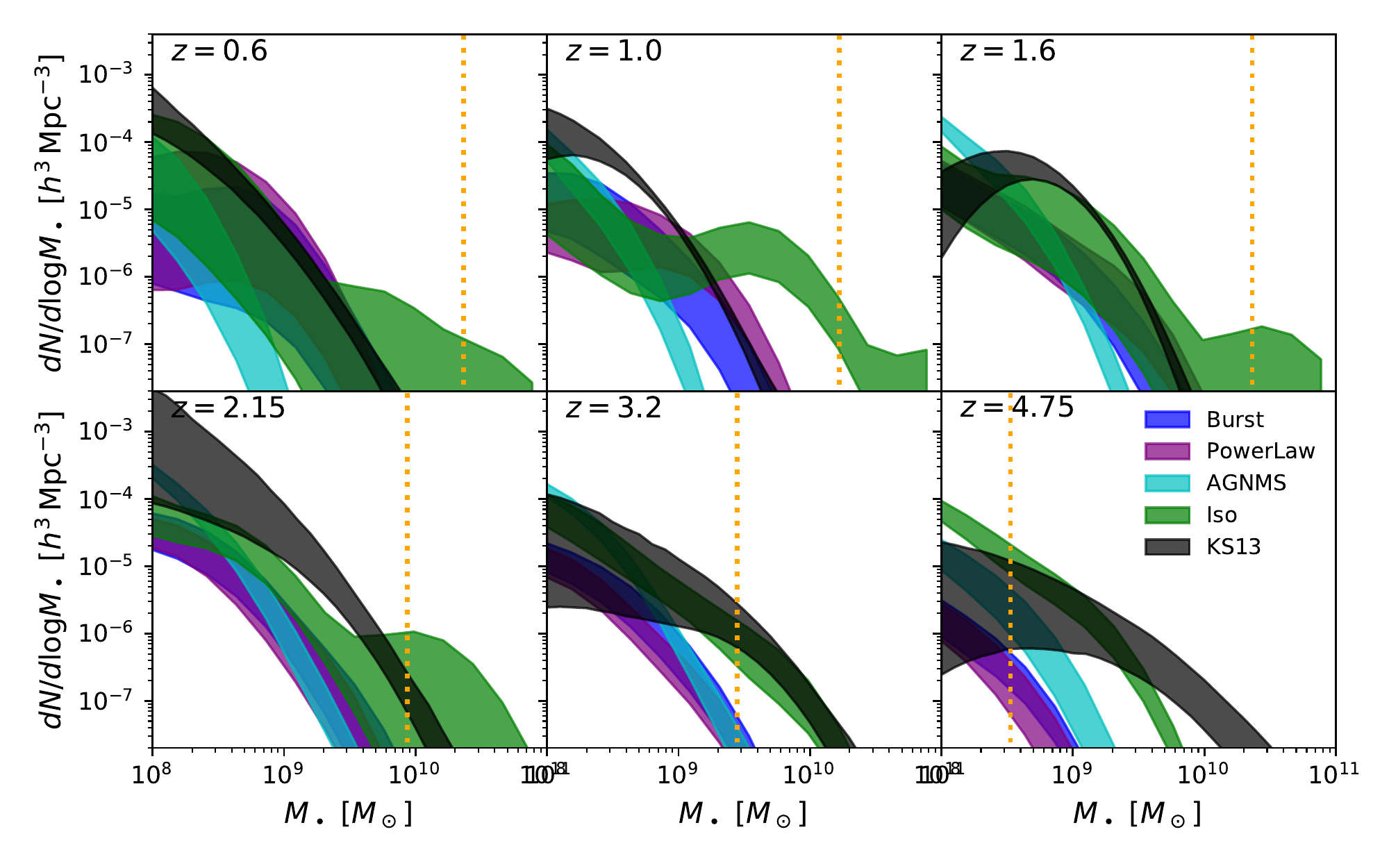} 
   \caption{Mass functions of BLQs compared to those inferred from SDSS \citep{Kelly&Shen2013}.  These predictions are sensitive to the details of the assumed Eddington ratio distribution.  Our models do not fare too poorly, but better agreement could be achieved by fine-tuning Eddington ratio distributions.  Note that the overall jump in normalisation in the data at $z=2.15$ is thought to be an artefact of the completeness correction in \citet{Kelly&Shen2013}.}
   \label{fig:mass_blq}
\end{figure*}

Using line-width relations, mass functions can also be inferred from broad-line quasars.  These mass functions probe the active and preferentially high-mass end of the population.  Since the probability of a SMBH being observed as a BLQ depends on both its luminosity and Eddington ratio, these mass functions are sensitive to the details of the underlying Eddington ratio distributions.  We do not tune anything in our model in an attempt to match these mass functions. 

Additional modelling is required to compute the probability of a SMBH being observed as a BLQ.  We follow \citet{Ueda+2014} to compute the probability that a SMBH is unobscured given its X-ray luminosity $L_x$ and $z$.  We assume the bolometric correction to the soft X-rays provided by the fitting function in \citet{Hopkins+2007}, reproduced below.  

\begin{align}
L_\mathrm{bol}/L_x = 10.83 \left( \frac{L_\mathrm{bol}}{10^{10} \ L_\odot} \right)^{0.28} + 6.08 \left( \frac{L_\mathrm{bol}}{10^{10} \ L_\odot} \right)^{-0.020}
\end{align}

Finally, we assume that all BLQs have an Eddington ratio of at least $10^{-2}$.  It is thought that broad-line regions do not exist below Eddington ratios of approximately this value \citep{Trump+2011}.  Our results are naturally sensitive to the value chosen for this threshold, as well as the power law parameters of the steady mode.

In Figure \ref{fig:mass_blq}, we compare our results with the inferred mass function of SDSS BLQs \citep{Kelly&Shen2013}.  Note that our completeness limits become more prohibitive as $z$ increases.  At $z=4.75$, these limits suggest that most of the progenitors of $>10^9 \ M_\odot$ SMBHs end up in halos with masses in excess of $10^{15} \ M_\odot$ by $z=0$.  None of Burst, PowerLaw, or Iso appear to be significantly preferred.  However, AGNMS clearly does not produce enough high-mass BLQs at low-redshift, since it requires the hosts of the most massive SMBHs to be star-forming for a BLQ to be observed.  Despite the addition of a steady mode, PowerLaw does not produce many more BLQs than Burst.  This is due to the minimum Eddington ratio threshold required for a SMBH to be a BLQ in this model.  Note that at $z=2.15$, where all of our models fall short, \citet{Kelly&Shen2013} report a discontinuity in the overall normalisation of their estimates, thought to be due to an error in their completeness correction.

Using a model most similar to our Iso variant, \citet{Natarajan&Volonteri2012} report a deficiency of high-mass BLQs in their highest redshift bin, $z=4.25$.  Up to our completeness limits, we do not report a deficiency of the same magnitude, perhaps due to accretion prescriptions that increase the probability of SMBHs being observed as BLQ.  In general, our models are not overly discrepant with observed BLQ mass functions.  Improving this agreement would require fine-tuning of Eddington ratio distributions.  In addition, higher-mass merger trees or alternative analytic techniques may be required to fully probe the high-mass, high-redshift population.

\section{Discussion}
\label{sec:discussion}

We have developed a SAM which is able to reproduce bolometric luminosity functions out to $z=6$ for accreting black holes while reaching the $M_\bullet-\sigma$ relation at $z=0$.  Here we highlight the significant consequences of choices made in modelling components and their subsequent impact on predictions.

\subsection{Three Channels of Black Hole Assembly}

In our model, SMBH mass assembly occurs through up to 3 different modes:  a merger-triggered and feedback-limited burst mode, a steady mode, and via direct SMBH mergers.  Here we discuss the regimes in which these three growth modes operate.

\subsubsection{The Burst Mode}
The burst mode is responsible for the bulk of SMBH growth.  It rapidly increases SMBH masses to match the $M_\bullet-\sigma$ relation.  This is especially important for high-redshift growth to explain the existence of $10^9 \ M_\odot$ SMBHs at $z \gtrsim 4$.  The amount of mass gained during the burst mode is a steep function of $\sigma$, and therefore the evolution of $M_\bullet$ is tied to the evolution of $\sigma$.  Recall that an emergent property of our $\sigma$ prescriptions is the assembly of the most massive SMBHs before less massive SMBHs.  At lower halo masses, a merger-triggered burst mode seems incapable of growing the low-mass SMBHs.  This is due to the fact that at low halo masses, $\sigma$, and therefore $M_\mathrm{cap}$, grows most rapidly at low-redshifts, when fewer mergers occur to trigger the burst mode.

\subsubsection{The Steady Mode}
The steady mode does not contribute to the growth of the most massive SMBHs, but it may dominate the assembly of the lowest mass SMBHs at low redshifts.  Among our four variants, only AGNMS is capable of assembling low-mass SMBHs without overgrowing the high-mass population.  This is because high-mass SMBHs finish assembling early in the universe, so additional growth modes that dominate at low-redshift must preferentially affect low masses.

 \subsubsection{SMBH Mergers}
SMBH mergers have the potential to contribute significantly to the assembly of the most massive SMBHs, where the occupation fraction is unity and the merger time-scales are short.  Consequently, SMBH mergers may increase the slope of the $M_\bullet-\sigma$ relation.  This growth channel proved so effective, in fact, that we were compelled to introduce the parameter $p_\mathrm{merge}$ to suppress this growth channel for low-redshift, high-mass SMBHs.

\begin{figure}
   \centering
   \includegraphics[width=0.45\textwidth]{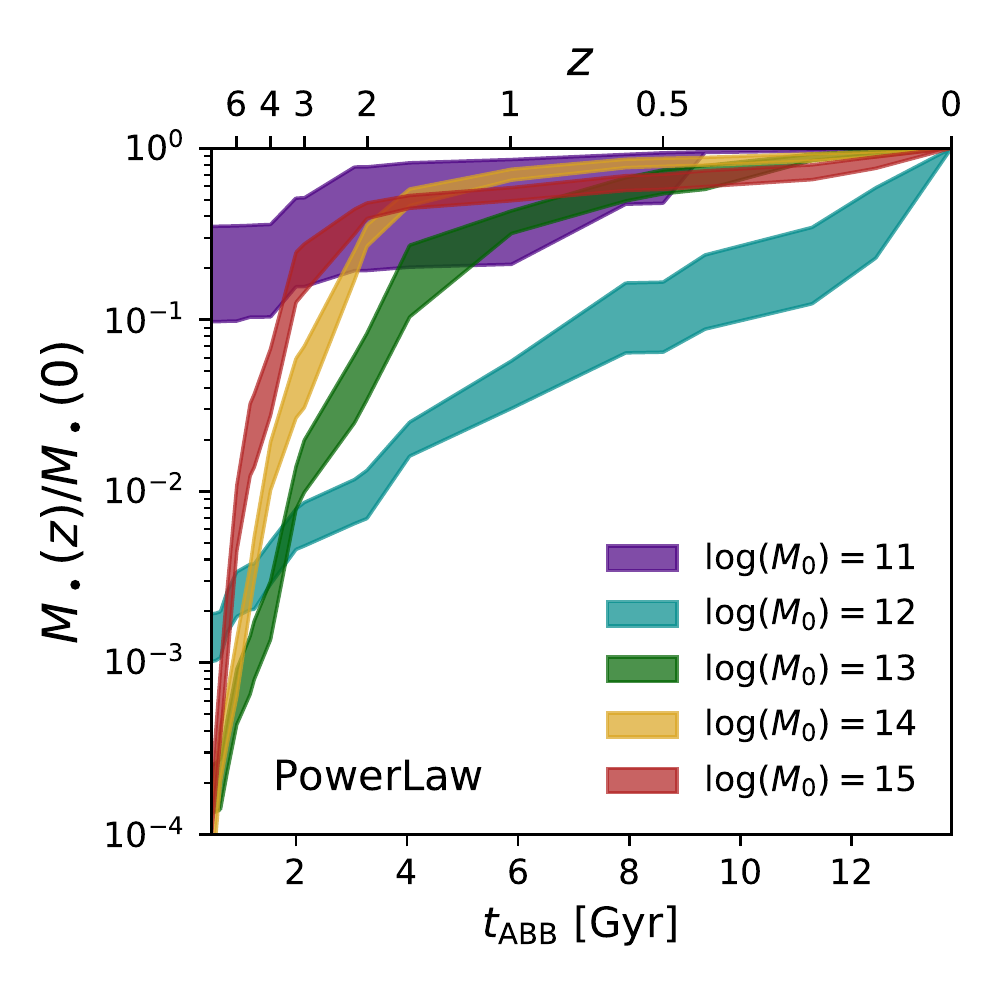} 
   \caption{The assembly history of SMBHs found in halos of different mass at $z=0$.  SMBHs in more massive halos assembled earlier in cosmic time.  In the lowest mass halos, SMBHs remain near their seed mass.  Such low-mass halos may preserve seeding information.}
   \label{fig:assemblyHistory}
\end{figure}

Figure \ref{fig:assemblyHistory} illustrates the assembly history of SMBHs in our PowerLaw model.  Here, the fraction of final mass assembled is plotted as a function of redshift for a range of final halo masses.  As described above, high-mass SMBHs assemble earlier than low-mass SMBHs.  Note that SMBHs hosted in $10^{11} \ M_\odot$ halos remain at masses comparable to their seed masses in this variant.  Consequently, these low-mass halos may have preserved information about the SMBH seed IMF \citep{Volonteri+2008,vanWassenhove+2010}.

\subsection{From Model to Observable}

Our model, and the SAM approach in general, contains a myriad of interconnected components and degenerate parameters that can be challenging to intuit.  Here, we attempt to provide an intuitive understanding of how each of our modelling components affects our observables, namely mass and luminosity functions.  This section is summarised in Table \ref{tab:modellingLessons}.

\begin{table*}
\centering
\begin{tabular}{llllll}
\cline{2-6}
                        & \multicolumn{2}{c}{Luminosity Functions}                                                                                                                    &  & \multicolumn{2}{c}{Overall Mass Functions}                                                                                                       \\
                        \hline
                        & High Redshift                                                       & Low Redshift                                                                          &  & High Redshift                                                      & Low Redshift                                                        \\
                        \hline
High Luminosity or Mass & \begin{tabular}[c]{@{}l@{}}Mass Cap\\ Burst Accretion\end{tabular}  & \begin{tabular}[c]{@{}l@{}}Steady Accretion\\ Mass Cap\end{tabular}                   &  & \begin{tabular}[c]{@{}l@{}}Mass Cap\\ Burst Accretion\end{tabular} & \begin{tabular}[c]{@{}l@{}}BH Mergers\\ Mass Cap\end{tabular}       \\
Low Luminosity or Mass  & \begin{tabular}[c]{@{}l@{}}Mass Cap\\ Steady Accretion\end{tabular} & \begin{tabular}[c]{@{}l@{}}Steady Accretion\\ Mass Cap\\ Burst Accretion\end{tabular} &  &  \begin{tabular}[c]{@{}l@{}}Mass Cap\\  \end{tabular}                                                           & \begin{tabular}[c]{@{}l@{}}Steady Accretion\\ Mass Cap\end{tabular} \\
\hline

                        &&&& \multicolumn{2}{c}{BLQ Mass Functions}                                                                                                                                                                                                                          \\
                        \hline
                        &&&& High Redshift                                                       & Low Redshift                                                                                                                                  \\
                        \hline
High Luminosity or Mass &&&& \begin{tabular}[c]{@{}l@{}}Mass Cap\\ Burst Accretion \end{tabular}  & \begin{tabular}[c]{@{}l@{}}Steady Accretion\\ Mass Cap\end{tabular}                          \\
Low Luminosity or Mass  &&&& \begin{tabular}[c]{@{}l@{}}N/A\\ \end{tabular} & \begin{tabular}[c]{@{}l@{}}N/A\\ \end{tabular}\\  \\
\hline

\end{tabular}
\caption{The dominant modelling components that influence each observable.  In each box, items are listed in order of importance.}
\label{tab:modellingLessons}
\end{table*}

\subsubsection{The Burst Mass Cap}

As described in equation \ref{eqn:mass_cap}, the mass cap determines the amount of mass gained during a major merger.  It determines the maximum mass of the SMBH hosted in a halo of a given $M_h$ and $z$.  The mass cap has two parameters, the normalisation $\alpha$ and the slope $\beta$.  These are tuned to provide reasonable matches to the local $M_\bullet-\sigma$ relation.  $\alpha$ and $\beta$ are degenerate with the parameters of the steady accretion mode and the black hole merger probability.  In order to obtain high-luminosity quasars at high redshift without overgrowing them by low redshift, $\alpha$ is tuned such that burst accretion is the dominant mode of SMBH growth for high masses.  We note that $\beta$ also has a minor effect in determining the slope of the luminosity function at high redshift.  Higher values of $\beta$ inhibit the burst mode in low-mass halos, resulting in a noticeably lower low-luminosity end.

\subsubsection{The Steady Mode}

The steady mode of SMBH growth dominates the luminosity function at redshifts $z < 2$.  We explore two possible implementations of this mode in PowerLaw and AGNMS.  Our power-law parameterisation of the steady mode has two parameters, the minimum Eddington ratio $f_{\mathrm{Edd},min}$ and the slope $dp/d\log f$.  These two parameters are tuned by-eye to match the $z=0.1$ luminosity function, and are assumed for simplicity not to vary as a function of either $z$ or $M_*$.  Consequently, as $z$ increases up to $2$, we obtain worse matches to the observed luminosity function.  The parameters chosen in our  PowerLaw variant allows high-mass SMBHs to contribute to the luminosity function without greatly increasing their mass.  However, the mass-independence of this mode results in under-massive SMBHs in halos with masses less than 
$\sim 10^{12} \ M_\odot$ at $z=0$.

Alternatively, our AGNMS variant ties the steady mode to star formation, which preferentially boosts the mass of SMBHs in low-mass halos.  In this model, the steady mode is able to grow the SMBHs which the burst mode does not due to the stochasticity of halo mergers.  However, this implementation does not produce any low-redshift, high-luminosity quasars. 

\subsubsection{Dynamics}

In our model, halos merge on the dynamical friction time-scale.  When a major merger occurs, SMBHs also have a chance to merge given by the free parameter $p_\mathrm{merge}$.  This single parameter is introduced to suppress mergers as a growth channel at the highest masses.  Cosmological simulations with more realistic treatments of dynamical friction suggest that there may be significant delays between halo merger and SMBH binary formation, on the order of Gyrs \citep[see recent results in][]{Tremmel+2017b}.  We also do not take into account the hardening of SMBH binaries, although gas dynamical friction, stellar scattering, and/or triaxiality are thought to rapidly evolve binaries to the gravitational wave regime.  Investigating the effects of realistic coalescence times will be the subject of future investigation.

\subsection{What constraints are we missing?}

Here, we comment on constraints from either observations or simulations that would be most helpful in advancing semi-analytic modelling.

\subsubsection{The AGN-Merger Connection}

In our model, mergers trigger the most voracious AGN, but not all AGN are triggered by mergers.  Mergers are responsible for triggering rapid growth periods, which in turn establish the slope of the $M_\bullet-\sigma$ relation.  Their frequency also regulates the duty cycle of these rapid growth periods, while their stochasticity contributes approximately 0.15 dex to the scatter in the $M_\bullet-\sigma$ relation.  In our model, mergers do not occur frequently enough at low redshift for the burst mode to consistently assemble SMBHs with masses less than $10^{7} \ M_\odot$.  One way to solve this problem would be for other processes besides mergers to trigger periods of rapid growth. Currently, the field is unsettled regarding the role of mergers in triggering AGN, and even if they do indeed trigger AGN, we are unaware of a comprehensive study that constrains the amount of mass gained during a merger in a realistic cosmological environment.  Advancements in both simulations and observations will be extremely useful to test this fundamental assumption in modelling. The definite test of whether major mergers drive the black hole growth process will be tested once the LISA experiment is launched and gravitational waves emitted from these myriad mergers are detected. 

\subsubsection{Long Term Growth Versus Short Term Variability}

We have found that a steady mode tracing the star formation rate is sufficient to explain the low-mass end of the local SMBH mass function, but the luminosity functions predicted with this mode are highly skewed toward low-luminosity.  On the other hand, a steady mode tailored to match the $z=0$ luminosity function fails to grow the population of low-mass SMBHs.  One possible solution to this quandary is that the growth rate over $\sim 10^7$ years may not match the instantaneous growth rate inferred when a luminosity is measured.

Variability on time-scales shorter than $10^6$ years, sometimes called AGN ``flickering,'' may be important to consider in future modelling \citep{Hickox+2014,Schawinski+2015}.  Observationally, there are few avenues available to directly probe these time-scales.  Thermal and ionisation echoes (``Voorwerpjes'') may contain the only direct constraints to variability on these time-scales \citep{Keel+2012,Schirmer+2013,Schirmer+2016}.

\subsubsection{SMBHs in Low-mass Hosts}

We have seen that various recipes for the the implementation of the steady mode of SMBH growth can have significantly different predictions for the low-mass end of the $M_\bullet-\sigma$ relation at $z=0$.  Additional statistics. more dynamical measurements of SMBHs in dwarf galaxies will help constrain this steady mode of growth.  It would be interesting if either the normalisation or scatter changed, and in this regard, it is intriguing that 
the $10^5 \ M_\bullet$ SMBH, one of the lowest inferred black hole masses, in RGG 118 lies on the $M_\bullet-\sigma$ relation\citep{Baldassare+2015}.

\section{Conclusion}
\label{sec:conclusion}

We have explored the assembly of SMBHs by developing a semi-analytic model to match the observed $M_\bullet-\sigma$ relation, mass functions, and luminosity functions with a variety of physical recipes.  We have explored 4 important variations of this model (i) comprising only merger-triggered Eddington-limited accretion; (ii) including a steady mode to match the local luminosity function; (iii) including a steady mode to match the local $M_\bullet-\sigma$ relation, and (iv) with a galaxy-halo-black hole mapping adopting the isothermal density profile and consequent velocity dispersion profile.  The most significant insights gained through this process are as follows:

\begin{itemize}
\item Isothermal spheres can perilously overestimate $\sigma$, and therefore $M_\bullet$. Previous works grossly overestimate the abundance of high-mass SMBHs at low redshift by setting  $\sigma = V_\mathrm{c,iso} / \sqrt{2}$ \citep{Natarajan&Volonteri2012}.  More careful derivations of $\sigma$ are necessary to match the local mass function.  
\item Galaxies with higher $\sigma$, and therefore $M_\bullet$, assemble earlier in the universe.  This arises from  the redshift-evolution of the stellar mass-halo mass relation and the mass-size relation, and this is enables matching the properties of high-luminosity, high-redshift AGN.
\item There are trade-offs matching $M_\bullet-\sigma$ versus matching the luminosity functions.  The steady accretion mode dominates for $z \lesssim 1$, simultaneously determining the growth of low-mass SMBHs and the luminosity function observed.  We find that a steady mode tuned to match the local luminosity function (PowerLaw) fails to grow most low-mass SMBHs to the $M_\bullet-\sigma$ relation, while a steady mode capable of growing these low-mass SMBH (AGNMS) fails to reproduce the luminosity functions at all redshift.  Perhaps the discrepancy stems from the fact that the growth of SMBHs on long time scales need not match the ``instantaneous'' luminosity observed today.
\item Intrinsic scatter that is not captured in the SAM cannot be neglected.  While the varied merger histories in our SAM produce approximately 0.15 dex intrinsic scatter, the data demand an additional 0.3 dex scatter.  This is necessary for producing objects of high mass and luminosity, as the number of density of SMBHs above $\sim 10^8 \ M_\odot$ is determined mostly by the number of lower mass SMBHs that scatter up.
\end{itemize} 

\section*{acknowledgements} Angelo Ricarte performed this research with the support from a Gruber Science Fellowship and from a Theoretical and Computational Astrophysics Network (TCAN) grant with award number 1332858 from the National Science Foundation (NSF) awarded to PN.  PN acknowledges support from the NSF TCAN  1332858. We thank Bhaskar Agarwal and Nico Cappelluti for helpful discussions, and Marta Volonteri, Michael Tremmel, and Rachel Somerville for comments on the manuscript.

\bibliography{ms}

\appendix

\section{Testing Our Velocity Dispersion Prescriptions}
\label{sec:mstar_sigma}

\begin{figure}
   \centering
   \includegraphics[width=0.45\textwidth]{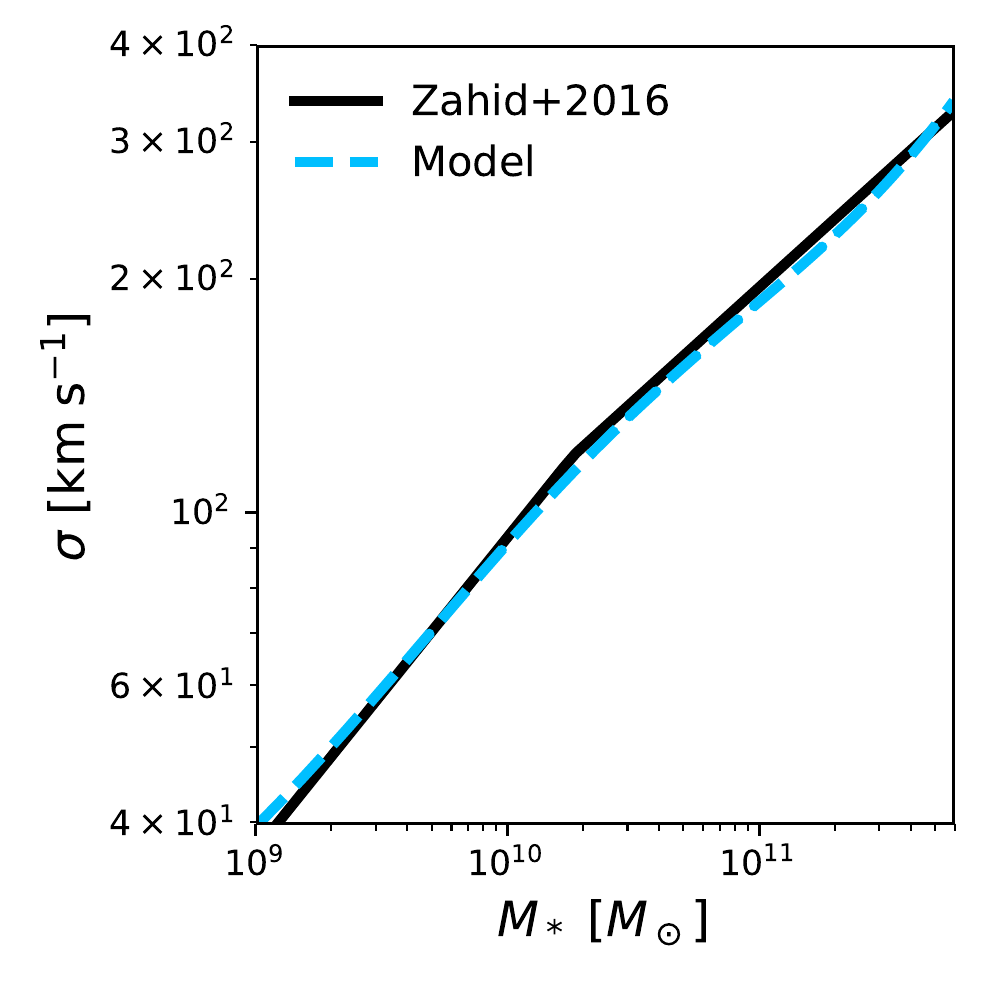}
   \caption{A comparison of our definition of $\sigma$ with that of local quiescent galaxies.  We find excellent agreement.}
   \label{fig:mstar_sigma}
\end{figure}

The velocity dispersion, $\sigma$, is used in our model to institute a cap on the accreted mass for black holes during the burst growth mode.  Here, we verify that we predict accurate values of $\sigma$ for local quiescent galaxies.  Figure \ref{fig:mstar_sigma} compares $\sigma(M_*)$ predicted through the various correlations exploited by our model (as shown in Fig.~4) and the local relation from SDSS \citep{Zahid+2016}.  Our model is in remarkable agreement with this relation, parametrized as a broken power law. 

\section{Computing Additional Scatter}
\label{sec:scatter}

Our model uses a series of scaling relations to forge the connection between halo mass and black hole mass.  Throughout this process, we expect to accumulate some amount of scatter that must be taken into account when constructing mass and luminosity functions.  Here, we describe the process by which we use the local measured SMBH mass function to estimate the amount of scatter that needs to be injected into our model.

The local SMBH mass function can be estimated from various scaling relations to indirectly acquire SMBH mass for a large number of SMBHs \citep[see][for a review]{Kelly&Merloni2012}.  Assuming that each SMBH and its host followed the relations in our model exactly, we can estimate the SMBH mass function analytically.  It is given by 

\begin{align}
\frac{dN}{d\log M_\bullet} = \frac{dN}{d\log M_h} \frac{d\log M_h}{d\log \sigma} \frac{d \log \sigma}{d \log M_\bullet}. \label{eqn:analyticMassFunction}
\end{align}

The first of these derivatives is the halo mass function.  We calculate the second of these derivatives numerically following the recipes chosen in our model.  The third derivative is the slope of the $M_\bullet-\sigma$ relation.  Since we must choose one, we select the relation obtained in \citet{Kormendy&Ho2013}, $\log (M_\bullet/M_\odot) = 8.49 + 4.38\log(\sigma / 200 \; \mathrm{km}\;\mathrm{s}^{-1})$.

\begin{figure}
   \centering
   \includegraphics[width=0.45\textwidth]{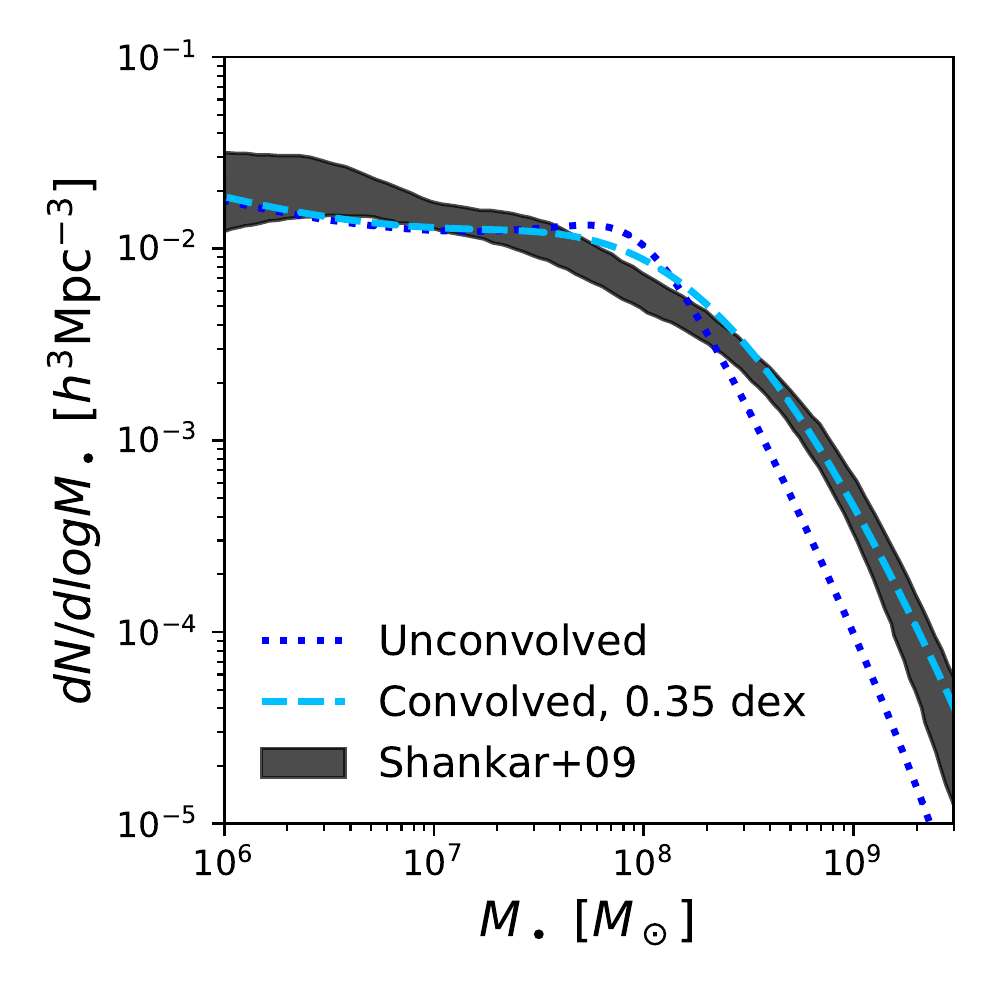}
   \caption{Analytic mass functions compared to the range of observational estimates \citep{Shankar+2009}, before and after convolution.  A total amount of injected scatter equal to 0.35 dex is required for the best agreement.}
   \label{fig:analyticMassFunctions}
\end{figure}

We then convolve this mass function with a log-normal kernel and vary its width by-eye until the observed SMBH mass function is matched.  The width of this kernel is then defined as the total scatter that needs to be included in our model.  Figure \ref{fig:analyticMassFunctions} displays the results of this exercise.  As shown, we arrive at a total scatter equal to 0.35 dex, which is reasonable as the $M_\bullet-\sigma$ relation itself has comparable or even greater scatter.  As shown in figure \ref{fig:msigma}, our model already generates some scatter due to a variety of merger histories.  We measure this intrinsic scatter at approximately 0.15 dex.  Therefore, we convolve all mass and luminosity functions generated by our model by a kernel of width 0.3 dex, which upon summing in quadrature with 0.15 dex yields the final desired 0.35 dex scatter.

Repeating this exercise for the Iso variant, for which $d\log M_h/d\log \sigma$ is different, we find there is no value of the intrinsic scatter for which the local mass function can be reproduced.  For self-consistency, we assume the same value of scatter for this variant.

\section{Estimating Black Hole Completeness}
\label{sec:completeness}

Due to computational constraints, the largest halo in our ensemble of merger trees has a mass of $10^{15} \ M_\odot$.  This places a limit on the maximum SMBH mass generated by our SAM.  These limits are shown in Figures \ref{fig:mass}, \ref{fig:lum}, and \ref{fig:mass_blq}.  Here, we outline how these limits are estimated.

The ``correct,'' unconvolved SMBH mass function at each redshift is calculated via equation \ref{eqn:analyticMassFunction}, using a halo mass function from HMFCALC \citep{Murray+2013}.  The SMBH mass function from our merger trees is then calculated using a halo mass function computed by binning the halos in the merger tree output files.  Each of these is convolved with a lognormal with a width of 0.3 dex to account for scatter in our scaling relations.

We then find the lowest SMBH mass for which the mass function from the merger tree is below that of HMFCALC by at least 50 per cent.  This is defined to be the highest mass for which the merger trees are complete, and the Eddington luminosity of this mass is the highest luminosity for which the merger trees are complete.  As shown in Figure \ref{fig:lum}, the most luminous quasars at $z=6$ are missing from our model; we predict that these would belong to the central galaxies of superclusters today.

\end{document}